\documentclass[twocolumn,aps,prd,showpacs,superscriptaddress,floats,floatfix,showkeys,notitlepage]{revtex4-1}

\usepackage{enumitem}
\usepackage{times}
\usepackage{dcolumn}
\usepackage{mathrsfs}
\usepackage{hyperref}
\usepackage{amsmath}
\usepackage{mathtools}
\usepackage{bbm}
\usepackage{bm}
\usepackage{amsfonts}
\usepackage{amssymb}
\usepackage{mathrsfs}
\usepackage{wasysym}
\usepackage{graphicx}
\usepackage[]{subfigure}
\usepackage{filecontents}
\usepackage[dvipsnames]{xcolor}
\usepackage{bbold}
\usepackage{lipsum}
\usepackage{graphicx}
\usepackage{subfigure}
\usepackage{palatino}
\usepackage{changes}
\usepackage{hyperref}
\hypersetup{colorlinks=true,linkcolor=blue,urlcolor=blue,citecolor=blue}
\usepackage[toc,page]{appendix}
\usepackage[normalem]{ulem}
\usepackage{adjustbox}
\usepackage{latexsym}
\usepackage{amsmath}
\usepackage{amssymb}
\usepackage{amsfonts}
\usepackage{times}
\usepackage{dcolumn}
\usepackage{bm}
\usepackage{tikz}
\usepackage{bigints}
\usepackage{array,tabularx,multirow}
\usepackage[dvipsnames]{xcolor}
\usepackage[tracking=true]{microtype}
\SetTracking{}{500}
\SetTracking{encoding={*}, shape=sc}{40}
\UseRawInputEncoding 
\allowdisplaybreaks

\usepackage{braket}
\usepackage{tensor}
\usepackage{slashed}
\usepackage{booktabs} 
\usepackage{float}

\begin{document} \sloppy

\title{Shadow and Weak Gravitational lensing of rotating traversable Wormhole in Non-homogeneous Plasma Space-time}

\author{Saurabh Kumar}
\email{ksaurabhkumar.712@gmail.com}
\affiliation{Department of Physics, Indian Institute of Technology, Guwahati 781039, India}

\author{Akhil Uniyal}
\email{akhil\_uniyal@sjtu.edu.cn}
\affiliation{Department of Physics, Indian Institute of Technology, Guwahati 781039, India}
\affiliation{Tsung-Dao Lee Institute, Shanghai Jiao Tong University, Shengrong Road 520, Shanghai, 201210, People’s Republic of China}

\author{Sayan Chakrabarti}
\email{sayan.chakrabarti@iitg.ac.in}
\affiliation{Department of Physics, Indian Institute of Technology, Guwahati 781039, India}

\date{\today}
\begin{abstract}
In this work, we have studied the behavior of null geodesics within a rotating wormhole space-time in non-magnetized pressure-less plasma. By focusing on the dispersion relation of the plasma and disregarding its direct gravitational effects, we examine how light rays traverse in the mentioned space-time. A key highlight of the work is the necessity of a specific plasma distribution profile to establish a generalized Carter's constant, shedding light on the importance of this parameter. Furthermore, we have derived analytical formulas to distinguish the shadow boundary across various plasma profiles, uncovering a fascinating trend of diminishing shadow size as plasma density increases. Intriguingly, certain limits of the plasma parameters result in the complete disappearance of the shadow. When calculating the deflection angle by a wormhole in plasma space-time, we observe a distinct pattern: the angle decreases as the plasma parameter rises in non-homogeneous plasma space-time, diverging from the behavior observed in homogeneous plasma space-time. Also, leveraging observational data from M$87^{\ast}$, we establish constraints on the throat radius. Furthermore, minimum shadow diameters provide valuable constraints for the radial and latitudinal plasma parameters. 

\end{abstract}

\maketitle

\date{\today}


\section{Introduction}\label{sec:intro}
The concept of a wormhole, a hypothetical structure in space-time that connects different regions in space-time, has been extensively explored since the early work of Einstein and Rosen \cite{Einstein1935}. Subsequent developments by Wheeler \cite{Wheeler1962} and the pioneering work of Morris and Thorne \cite{Morris1988} on traversable static wormholes further fueled interest in these intriguing cosmic constructs. In a later work, Teo extended this concept to include rotation in the wormhole geometry \cite{Teo1998}. The existence of wormholes challenges energy conditions and requires the presence of exotic matter within the throat \cite{Morris1988}. Consequently, their plausibility has been a subject of debate. Various proposals, such as the existence of a thin layer of negative energy density inside the throat \cite{Visser_2003} or the incorporation of modified gravity theories \cite{Maeda2008, Lobo2009} have been put forth to address these challenges. Given the potential formation of wormholes in the early universe and their existence subject to specific conditions, it is essential to investigate them further and discern their unique characteristics from other compact objects. In the literature, there are different methods that exist in order to get the axisymmetric wormhole solution. One such example is using the Ehlers transformations \cite{Cisterna:2023uqf}.

It is believed that the center of the galaxies including ours contains supermassive black holes however the existence of the black hole can only be justified by the existence of the event horizon. Therefore a number of tests have been proposed in order to confirm the presence of an event horizon in such compact objects \cite{Broderick:2015tda, Broderick:2009ph, Narayan:2008bv}. In spite of all these proofs, a shred of conclusive evidence is still lacking \cite{Abramowicz:2002vt}. It is worth mentioning that the existence of the event horizon along with the set of unstable light rings commonly known as the photon sphere in the exterior region of the compact object form the shadow of the object with the help of the radiation coming from the accretion disk around it \cite{Takahashi:2004xh}. Therefore, since the publication of the image of the M$87^{\ast}$ supermassive black hole \cite{EHT2019_1} and supermassive compact object at the center of our own galaxy known as Sagittarius $A^{\ast}$ (Sgr $A^{\ast}$) \cite{EventHorizonTelescope:2022wkp} by the Event Horizon Telescope (EHT), there has been extensive discussion among researchers regarding the nature of the object captured in the image including one of the papers by EHT \cite{EventHorizonTelescope:2022xqj}. However, first Synge \cite{Synge:1966okc} and Luminet \cite{Luminet:1979nyg} studied the Schwarzchild black hole shadow, and thereafter Bardeen \cite{1973blho.conf.....D} looked into the shadow of Kerr black hole. Consequently, the shadow in different geometrical backgrounds was studied in detail such as the Reissner-Nordstr\"{o}m (RN) black hole \cite{Zakharov:2014lqa}, Kerr-Newman black hole \cite{Takahashi:2005hy}, rotating regular black hole \cite{PhysRevD.97.064021}, Kerr black hole with scalar hair \cite{Cunha:2015yba}, regular black hole \cite{Li:2013jra, Abdujabbarov:2016hnw, Amir:2016cen}, Einstein-dilaton-Gauss-Bonnet black holes \cite{Cunha:2016wzk}, Horava-Lifshitz black holes \cite{Atamurotov:2013dpa}, non-Kerr black holes \cite{Atamurotov:2013sca, Wang:2017hjl}, higher-dimensional black holes \cite{Papnoi:2014aaa, Uniyal:2022xnq, Abdujabbarov:2015rqa}, black holes in theories of Non-linear electrodynamics \cite{Okyay:2021nnh, Uniyal:2022vdu, Uniyal:2023inx}, black holes in loop quantum gravity \cite{Liu:2020ola,Devi:2021ctm, Afrin:2022ztr}, black holes in the expanding Universe \cite{Perlick:2018iye, Roy:2020dyy}, black holes in the presence of plasma \cite{Atamurotov:2015nra, Abdujabbarov:2015pqp} etc. to name a few. It is important to understand that while the boundary of the shadow is only determined by the underlying space-time metric since it is formed only by the observed apparent shape of the photon sphere by the distant observer \cite{Perlick:2021aok}, the intensity map of the image is influenced by the accretion process around the compact object. Therefore, it is important to note that the presence of a shadow or a photon ring does not provide conclusive evidence that the object is a black hole. This also has been shown in the recent simulations by using the general relativistic magnetohydrodynamical (GRMHD) and general relativistic radiative transfer (GRRT) calculations that distinguishing the shadow image of the Kerr black hole and non-rotating dilaton black hole is almost impossible within the present observations \cite{Mizuno:2018lxz, Roder:2023oqa}. In support of the argument, a number of other compact objects have been studied where it has been shown that the horizonless compact object such as naked singularities \cite{Bambi:2008jg, Ortiz:2015rma, Shaikh:2018lcc}, a hard surface \cite{Broderick:2005xa}, and non-rotating wormholes \cite{Bambi:2013nla, Ohgami:2015nra, Azreg-Ainou:2014dwa} and rotating traversable wormhole \cite{Nedkova:2013msa} can also cast the similar shadows. Along with this wormhole shadow also has been discussed in arbitrary metric theory of gravity with parameterizing the wormhole space-time \cite{Bronnikov:2021liv}.

Previous studies have extensively investigated the shadows of wormholes, discussing their similarities with the shadow of Kerr black hole \cite{Nedkova:2013msa, abdujabbarov2016shadow, Shaikh2018}. However, one crucial aspect that has been overlooked in these studies is the presence of plasma and its effects on wormhole shadows. The effects of plasma on the shadows of rotating wormholes space-time have been explored in \cite{abdujabbarov2016shadow}, however, this study did not consider the contribution of the wormhole throat \cite{Shaikh2018}. Therefore, in this work, we will be studying the Teo class of rotating wormholes \cite{Teo:1998dp} in the presence of the plasma and will be taking care of the contribution coming from the throat. Other than the shadow, a lot of studies were performed on gravitational lensing for the wormhole without the plasma medium \cite{Cramer:1994qj, Nandi:2006ds, Nakajima:2012pu, Tsukamoto:2012xs, Tsukamoto:2016zdu, Jusufi:2017mav} and with the plasma medium \cite{Adam2015, Bisnovatyi-Kogan_2010}. Furthermore, investigations into weak lensing in plasma space-time have not been limited to compact objects alone, as some researchers have employed galaxy models to study the effects of non-uniform plasma, revealing an increasing impact on the deflection angle \cite{bisnovatyikogan2015gravitational,farruh2021}.

Our goal is to derive analytical expressions for the shadow boundary of the rotating wormhole in plasma-filled space-time for the observer situated at infinity, similar to Bardeen's calculation of the Kerr black hole shadow \cite{bardeen}. It is known that including the plasma potential in the Hamiltonian can affect the existence of Carter's constant, so one crucial aspect of our work will be to find the necessary condition for the existence of Carter's constant \cite{carter1968global}. Such a condition has also been pointed out for Kerr black hole and also for generalized axis-symmetric static-spacetime \cite{Perlick:2017fio, Bezdekova:2022gib}. Another aim is to derive the deflection angle by a wormhole in homogeneous and non-homogeneous plasma space-time and analyze their impacts on the deflection angle. At last, our final goal will be to constrain the wormhole and plasma parameters using the EHT results of black hole shadows at the center of M87*. A similar approach and calculations have been taken in \cite{Rahaman_2021}.

The paper is structured as follows: In Section II, we provide an overview of the Hamiltonian formalism for null geodesics in plasma space-time and discuss the necessary conditions for the existence of light rays in the outer communication of the Teo wormhole plasma space-time. Section III focuses on determining the specific forms of plasma profiles that satisfy the condition for the existence of Carter's constant. We derive the expressions for the null geodesic equations in this context. In Section IV, we delve into the role of the contribution of the wormhole throat to the wormhole shadow. We discuss the significance of the throat and derive the expressions for the celestial coordinates of the shadow boundary in generalized plasma space-time. Moving on to Section V, we explore specific plasma profiles that fulfill the separability condition. We also present a comparison of the shadows for various plasma densities of different plasma profiles. Section VI is dedicated to the calculation of the deflection angle of a rotating wormhole in weak field approximation for both homogeneous and non-homogeneous plasma space-time. Finally, in Section VII, we attempt to constrain the plasma parameters and throat size of the Teo wormhole using the observational data from M87*. Throughout the paper, we have considered units such that $\hbar=G=c=M=1$ and our choice of signature is (-,+,+,+).

\section{HAMILTON FORMALISM FOR LIGHT RAYS IN A PLASMA space-time}
The Hamiltonian describing the light ray traveling in non-magnetized pressureless plasma is given as \cite{Perlick:2017fio}
\begin{equation}
    H(x, p)=\frac{1}{2}\left(g^{\alpha \beta}(x) p_\alpha p_\beta+\omega_P(x)^2\right),
\label{2.1}
\end{equation}
here $g^{\mu \nu}$ are the contravariant components of the metric tensor and $\omega_P$ represents the plasma electron frequency, which is defined as,
\begin{equation}
    \omega_P(x)^2=\frac{4 \pi e^2}{m} N(x) ,
\label{2.2}
\end{equation}
where $m$ and $e$ are the mass and charge of the electron respectively while $N(x)$ defines the electron density distribution. Here $x$ represents the space-time coordinates ($t,r,\theta,\phi$) while $p$ represents the momentum coordinates ($p_t,p_r,p_\theta,p_\phi$) for the light ray. Please note that the plasma frequency ($\omega_P$) and the photon frequency ($\omega$) is related by a general form,
\begin{equation}
    n(x, \omega(x))^2=1-\frac{\omega_P(x)^2}{\omega(x)^2},
\label{2.3}
\end{equation}
where $n$ is known as the refractive index, and it must be greater than 0 so that the light rays reach the observer \cite{Perlick:2017fio}. Since the light rays reaching the observer are gravitationally redshifted, the observed redshifted frequency can be expressed in terms of the known constant of motion $p_t$ as
\begin{equation}
    \omega(x)=\frac{p_t}{\sqrt{-g_{tt}(x)}}.
\label{2.4}
\end{equation}
The necessary and sufficient condition for the existence of a light ray with a constant of motion $p_t$ is derived for the Kerr black hole by Volker et. al \cite{Perlick:2017fio} and based on a similar approach, a similar condition for generalized rotating metric can be given by 
\begin{equation}
    p_t^2 > g_{tt}(x) \omega_P(x)^2.
\label{2.7}
\end{equation}
Finally, the geodesics equations can be derived using Hamilton's equations, which are given as,
\begin{equation}
    \dot{p}_\alpha=-\frac{\partial H}{\partial x^\alpha}, ~ \dot{x}^\alpha=\frac{\partial H}{\partial p_\alpha}.
\label{2.6}
\end{equation}
For our analysis we have considered a stationary, axisymmetric rotating metric for  Teo class traversable wormhole in the Boyer-Lindquist coordinates as \cite{Teo1998},
\begin{align}
\begin{split}
ds^2 = & -N(r)^2 dt^2 + \left(1 - \frac{b_0(r)}{r}\right)^{-1} dr^2 \\
& + r^2 K(r)^2 \left(d\theta^2 + \sin^2\theta \left(d\phi - \omega_T(r) dt\right)^2\right),
\end{split}
\label{2.8}
\end{align}
where $r \geq r_0 $, $r_0$ is the throat radius of the wormhole. $N$, $b_0$ are known as the redshift factor and shape function respectively, $K$ determines the proper radial distance which is given by $R=rK$ and $\omega_T$ is the measure for the angular velocity of the wormhole. $N$, $b_0$, $K$ and $\omega_T$ are in general the functions of radial ($r$) and polar ($\theta$) coordinates. For simplicity, in this work, we have only considered $r$ dependency. Since the wormhole does not contain an event horizon, the metric component $N(r)$ should be considered finite and non-zero throughout space-time.
The shape function ($b_0$) must satisfy the conditions $\left.\partial_\theta b_0\right|_{r=r_0}=0,\left.\partial_r b_0\right|_{r=r_0}<1$ and $b_0 \leq r$ \cite{Morris1988} in order to have the geometry of a wormhole. The shape function also gives information about the mass of the wormhole \cite{Visser1996} and the estimation of mass as $M=r_0/2$ is derived in great detail by Shaikh et. al \cite{Shaikh2018}. In this work, we have considered the following form of the metric functions in order to get the traversable wormhole \cite{Teo1998, Nedkova2013}.
\begin{equation}
\begin{aligned}
& N=\exp \left[-\frac{r_0}{r}\right], \quad b_0(r)=r_0=2M, \\
& K=1, \quad \omega_T=\frac{2 J}{r^3},
\end{aligned}
\label{2.9}
\end{equation}
where $J$ is the angular momentum of the wormhole and $M$ is the mass of the wormhole \cite{Shaikh2018, Visser1996}. We have used spin parameter of wormhole, a=$J/M^2$ which defines its rotation rate.

\section{SEPARABILITY OF HAMILTON-JACOBI EQUATION FOR NULL GEODESICS IN PLASMA ON TEO WORMHOLE SPACE-TIME}
The geodesic motion in rotating space-time enables two constants of motion — the angular momentum of the particle about the axis of symmetry $p_{\phi}$ and its energy $p_t$ due to the axisymmetric and stationary symmetries of the space-time. However, Carter et al. \cite{carter1968global} showed that the geodesics in Kerr metric possess another constant of motion that governs the motion of geodesics in the latitudinal direction. Since the Kerr metric represents the rotating black hole space-time, Carter's constant should also exist in the rotating wormhole. This constant can be found using the method of separation of variables. Therefore, let's consider the Hamiltonian for the null geodesics as,\\
\begin{equation} 
   H\left(x,\frac{\partial S}{\partial x^\alpha} \right)=  \frac{1}{2} g^{\alpha \beta} (x)\frac{\partial S}{\partial x^\alpha} \frac{\partial S}{\partial x^\beta}+\frac{1}{2}\omega_P(x)^2=0,
\label{3.1}
\end{equation}\\
with the separation ansatz
\begin{equation} 
    S(t,r,\theta,\phi) =  p_t t + p_{\phi} \phi + S_r(r) + S_\theta (\theta),
\label{3.2}
\end{equation}
where $S_r(r)$ and $S_\theta (\theta)$ are functions of $r$ and $\theta$ coordinates respectively. Now substituting Eq. \ref{3.2} into Eq. \ref{3.1} will give,
\begin{align}
    \begin{split}
        \frac{1}{2} g^{t t}\left(\partial_{t} S\right)^{2}+ g^{\phi t} \left(\partial_{t} S\right) \left(\partial_{\phi} S\right)+\frac{1}{2} g^{r r}\left(\partial_{r} S\right)^{2}\\
+\frac{1}{2} g^{\theta \theta}\left(\partial_{\theta} S\right)^{2}
+\frac{1}{2} g^{\phi \phi}\left(\partial_{\phi} S\right)^{2}+\frac{1}{2} \omega _P ^2=0.
    \end{split}
\label{3.3}
\end{align}
Now considering $p_r = \partial_{r} S$ and $p_\theta = \partial_{\theta} S$ and solving the above equation for Teo rotating wormhole space-time (Eq. \ref{2.8}) will give,
\begin{align}
\begin{split}
    -\frac{1}{N^2} p_t^2-2 \frac{\omega_T}{N^2} p_t p_\phi+\left(1-\frac{b_0}{r}\right) p_r^2+\frac{1}{r^2K^2} p_\theta^2\\-\left(\frac{\omega_T^2}{N^2}-\frac{1}{r^2K^2 \sin ^2 \theta}\right) p_\phi^2 + \omega _P ^2 =0.
    \end{split}
\label{3.4}
\end{align}
Since we are considering plasma frequency which depends on both radial ($r$) and polar ($\theta$) coordinates, the above equation is only separable if the general form of plasma frequency is considered as,
\begin{equation}
    \omega _P (r,\theta)^2 = \frac{\Omega _r(r)+\Omega _{\theta}(\theta)}{r^2K^2},
\label{3.5}
\end{equation}
where $\Omega_r(r)$ and $\Omega_\theta(\theta)$ are $r$ and $\theta$ dependent functions respectively. Therefore, Eq. \ref{3.4} can be rearranged as,
\begin{align}
\begin{split}
    \underbrace{-\frac{r^2K^2}{N^2}\left(p_t+\omega_T p_\phi\right)^2+r^2K^2\left(1-\frac{b_0}{r}\right) p_r^2+\Omega _r(r)}_{f_r(r)} \\=
     \underbrace{-p_\theta^2-\frac{p_\phi^2}{\sin ^2 \theta}-\Omega _{\theta}(\theta)}_{f_{\theta}(\theta)} ,
     \end{split}
\label{3.6}
\end{align}
here expressions $f_r(r)$ and $f_{\theta}(\theta)$ are only the function of $r$ and $\theta$ respectively and therefore can be considered as a constant since they are now separated by equality. This constant is known as Carter's constant and can be written as,
\begin{equation}
    f_r(r)=f_{\theta}(\theta) = - Q.
\label{3.7}
\end{equation}
Therefore, by using these three constants of motion $p_t$, $p_{\phi}$, and $Q$, one can write the impact parameters such as \cite{Nedkova2013},
\begin{equation}
    \eta=\frac{L}{\omega_o}, ~ \xi=\frac{Q}{\omega_o^2},
\label{3.8}
\end{equation}
where we have considered $p_t=-\omega_o$ and $p_\phi=L$. Now solving for geodesics using Hamilton's Eqs. (\ref{2.6}) for $x^\mu=t, \phi$, we get:
\begin{equation}  
   \dot{t} =\frac{1}{N^{2}}\left(1-\eta \omega_{T}  \right),
\label{3.9}
\end{equation}

\begin{equation}
        \dot{\phi} =\frac{1}{N^2}\left(\omega_T\left(1-\eta \omega_T \right)+\eta \frac{N^2}{r^2K^2 \sin ^2 \theta} \right).
\label{3.10}
\end{equation}
By calculating the expressions for $p_r$ and $p_\theta$ using Eq. \ref{3.6},
\begin{equation}
   p_r= \pm \frac{1}{N\sqrt{1-\frac{b_0}{r}}}\sqrt{(1-\eta\omega_T)^2-\frac{N^2}{r^2K^2} \left(\xi+\frac{\Omega _r}{\omega_o^2}\right)},
\label{3.11}
\end{equation}
\begin{equation}
    p_{\theta} =\pm\sqrt{ \xi -\frac{\eta^2}{\sin{\theta}^2}-\frac{\Omega _{\theta}}{\omega_o^2}},
\label{3.12}
\end{equation}
we can calculate the remaining two geodesic Eqs. by solving Eqs \ref{2.6} for $x^\mu=r, \theta$ and we get,
\begin{equation}
    \dot{r} = \pm \frac{\sqrt{1-\frac{b_0}{r}}}{N}\sqrt{R(r)},
\label{3.13}
\end{equation}
\begin{equation}
    \dot{\theta} = \pm\frac{1}{r^2K^2}\sqrt{\Theta(\theta) },
\label{3.14}
\end{equation}
where $R(r)$ and $\Theta(\theta)$ are expressed as,
\begin{equation}
    R(r) = (1-\eta \omega_T)^2-\frac{N^2}{r^2K^2}\left(\xi+\frac{\Omega _r}{\omega_o^2}\right),
\label{3.15}
\end{equation}
\begin{equation}
    \Theta(\theta)=\xi -\frac{\Omega_{\theta}}{\omega_o^2}-\frac{\eta^2}{\sin{\theta}^2}.
\label{3.16}
\end{equation}
Since the shadow is formed due to last photon rings which are unstable in nature, so in order to have unstable spherical orbits, null rays must satisfy the following criteria \cite{Perlick:2017fio},
\begin{equation}
    \Theta(\theta)\geq 0 ~,~~R''(r) >0.
\label{3.17}
\end{equation}
where the first condition ensures the existence of spherical orbits around the wormhole while the second condition is imposed to get unstable spherical orbits. A similar calculation for the general axially symmetric stationary space-time case can also be found in \cite{Bezdekova:2022gib}.

\section{SHADOW OF TEO WORMHOLE IN PLASMA space-time}
Since photon orbits offer valuable insights into the optical appearance of wormholes, it would be insightful to study the boundary of the last photon ring in plasma space-time. In a non-rotating space-time, these orbits occur within the equatorial plane due to the spherical symmetry of the wormhole. However, in the case of rotating space-time, photon trajectories cross the equatorial plane repeatedly \cite{Teo2003}. The Carter's constant which remains conserved in the latitudinal direction is crucial in order to determine the spherical orbits.

The primary objective is to identify the last photon orbits that distinguish between light rays moving outward and those moving inward. To accomplish this, we rely on the determination of critical orbits characterized by their impact parameters: $\eta$ and $\xi$. These parameters play a pivotal role in delineating the boundary of the shadow cast by the wormhole. Remarkably, the last photon orbits correspond to the most unstable circular orbits, featuring the maximum value of the effective potential, $V_{eff}$. Well-established criteria can be applied to identify this unstable circular photon orbits \cite{Frolov1998}:
\begin{equation}
    V_{eff}(r_c)=0, \quad {V{'}_{eff}}(r_c)=0,
\label{4.1}
\end{equation}
where $r_c$ denotes the critical photon orbits and prime denotes the derivative with respect to $r$. The geodesic equation, Eq. \ref{3.13} can be expressed as,
\begin{equation}
    \dot{r}^2+V_{eff}=0,
\label{4.2}
\end{equation}
where
\begin{equation}
    V_{eff} =-\frac{1-b_0/r}{N^2}\Bigg[(1- \eta \omega_T)^2 - \frac{N^2}{r^2K^2} \left( \xi + \frac{\Omega_r}{\omega_o^2} \right) \Bigg],
\label{4.3}
\end{equation}
Hence, calculating the impact parameter with the help of Eqs. \ref{4.1}, we get,
\begin{equation}
\xi = \left. \left[\frac{r^2K^2}{N^2}(1-\eta\omega_T )^2-\frac{\Omega_r}{\omega_o^2}\right]\right|_{r=r_c},
\label{4.4}
\end{equation}
\begin{equation}
     \eta = \left.\frac{B-\sqrt{B^2-4AC}}{2A}\right|_{r=r_c},
\label{4.5}
\end{equation}
\begin{figure*}
\subfigure(a){\includegraphics[width = 2.8in, height=2.6in]{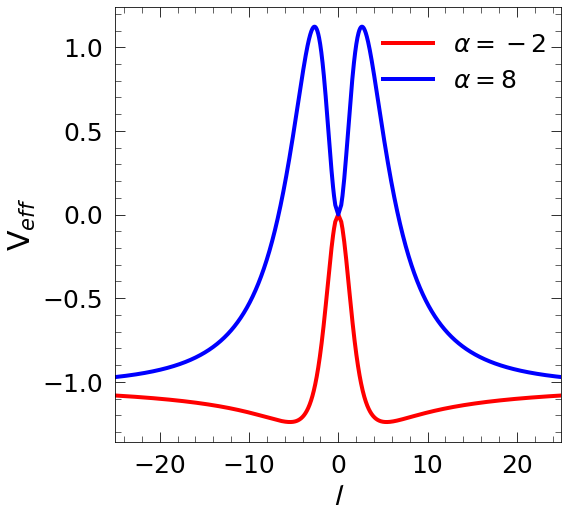}}
\subfigure(b){\includegraphics[width = 2.8in, height=2.6in]{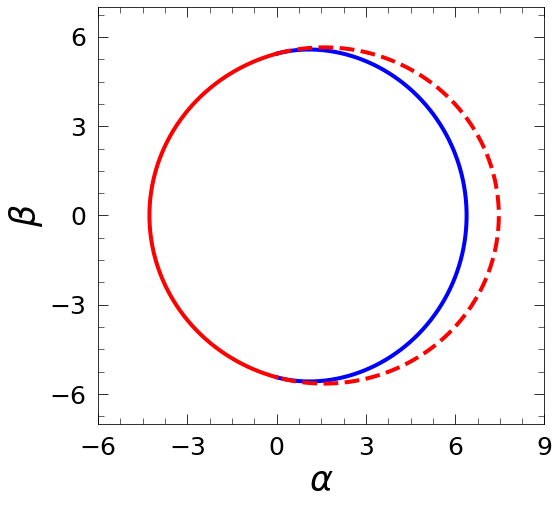}}
\caption{(a) Effective potential of the Teo wormhole with $\beta=0$ and $\alpha=-2$ (red), $8$ (blue). These plots reveal that for positive values of alpha, there is an extremum outside the throat, while for negative values of alpha, the extremum is located at the throat. This distinction helps us understand the respective contributions of the potential in the formation of the shadow. [See text for more details.] (b) Wormhole shadow (blue) due to maximum potential outside the throat and (red) due to unstable orbits at the throat. Here $l$ is the proper radial distance given by $l(r)=\pm \int_{r_0}^r \frac{dr}{\sqrt{1-\frac{r_0}{r}}}$ }
\label{fig:1}
\end{figure*}
where
\begin{align}
         A &= \omega_{T} \omega_{T}^{\prime}-\omega_{T}^{2} \Sigma,\\
    B &= \omega_{T}^{\prime}-2 \omega_{T} \Sigma, \\
    C &= \Bigg(\frac{\Omega_{r}}{\omega_{o}^{2}} \frac{N^{2}}{r^2K^2}-1\Bigg) \Sigma+\Delta,\\
         \Delta &= \frac{1}{2}\frac{d}{dr}\left(\frac{N^2}{r^2K^2}\frac{\Omega_r}{\omega_o^2}\right),\\
         \Sigma&=\frac{1}{2} \frac{d}{d r}\left(\ln \frac{N^{2}}{r^{2} K^2}\right).
\label{4.6}
\end{align}
Since we have discussed that these critical orbits are crucial in determining the last photon rings, therefore, $\eta$ and $\xi$ can completely determine the boundary of the shadow however in order to look for the shadow in the observer's sky, we have used the following definitions of celestial coordinates \cite{Vázquez_Esteban_2004}:
\begin{align}
    \alpha &=\lim _{r \rightarrow \infty}\left(-r^2 \sin \theta \frac{d \phi}{d r}\right),\label{4.7a}\\
    \beta &=\lim _{r \rightarrow \infty} r^2 \frac{d \theta}{d r},
\label{4.7b}
\end{align}
and these celestial coordinates can be calculated with the help of impact parameters, $\eta$, and $\xi$ by following the geodesic Eqs. derived in section III and given as,
\begin{align}
    \alpha &= -\frac{\eta}{ \sin{\theta}},   \label{4.8a}\\
    \beta &= \sqrt{\xi-\frac{\eta^2}{\sin^2{\theta}}-\frac{\Omega_{\theta}}{\omega_o^2}}.
\label{4.8b}
\end{align}
These expressions are not valid for calculating the celestial coordinates in homogeneous plasma space-time which will be discussed in the next section.

Another crucial factor to consider in the case of the wormhole shadow is the existence of the extremum potential at the throat of the wormhole. It becomes apparent from Eq. \ref{4.3} that the effective potential becomes zero at the throat when $r=r_0$. This implies that stable or unstable spherical orbits may exist depending on the sign of the second derivative of the effective potential (${V{''}_{eff}}(r_0)$). To gain insights into the formation of the shadow, Fig. \ref{fig:1}(a) showcases the effective potential for values of $\alpha$ equal to $-2$ (red), $8$ (blue), and $\beta$ equal to $0$. It has been observed that for positive values of $\alpha$ the potential exhibit two extrema however unstable orbits are located outside the throat. Consequently, the contribution to the shadow is solely derived from the outer region. On the other hand, for negative values of $\alpha$, only one extremum is present at the throat. Please note that Fig. \ref{fig:1}(a) is not solely responsible for the formation of complete shadow as shown in Fig. \ref{fig:1}(b). The potential is for illustration purposes for showing the existence of extrema at the throat and outside the throat. It is noteworthy to point out that previous studies on the formation of wormhole shadows in plasma space-time failed to account for the contribution of the throat potential, despite the throat contributions being highlighted in rotating wormholes \cite{Shaikh2018}.
\begin{figure*}
\subfigure(a){\includegraphics[width = 2.8in, height=2.7in]{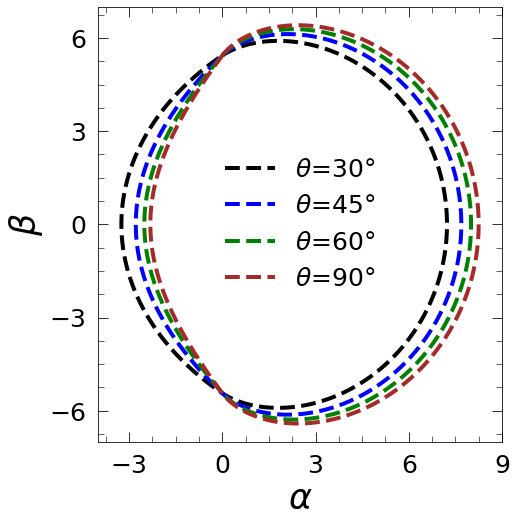}}
\subfigure(b){\includegraphics[width = 2.8in, height=2.7in]{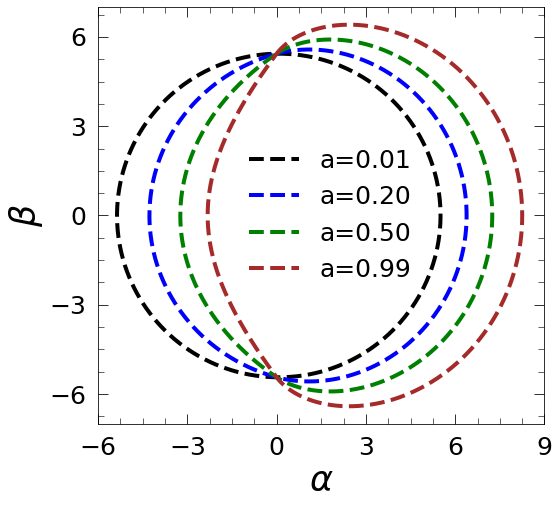}}
\caption{Wormhole shadows in a vacuum with throat radius, $r_0=2$ (a) for various inclinations angles with spin parameter, $a=0.99$ and (b) for various spin parameters with inclination angle, $\theta=90^\circ$.}
\label{fig:2}
\end{figure*}
In our quest to understand the intricate interplay of factors contributing to shadow formation, we delve into the analysis of spherical photon orbits that satisfy constraints Eqs. \ref{3.17}. We solve the Eq. \ref{4.8b} for $\beta=0$ along with satisfying the constraint equations to find out the minimum ($r_{min}$) and maximum ($r_{max}$) radius for the shadow formation. Therefore, the shadow will consist of the orbit formed by ($r_{min}, r_{max}$). However, it turns out that sometimes $r_{min}$ can be less than the throat radius $r_0$. In such cases, the shadow will be formed by the orbits consisting of ($r_0, r_{max}$), highlighting the contribution of the throat. This nuanced differentiation ensures that we gain a comprehensive understanding of the shadow formation mechanism, considering the varying contributions from different regions of the wormhole's geometry.

Since at the throat, the potential vanishes which also corresponds to the extremum of the potential (see Fig. \ref{fig:1}(a)) therefore using Eq. \ref{4.3},
\begin{equation}
    \left.\left[\left(1-\eta \omega_{T}\right)^{2}-\frac{N^{2}}{r^{2} K^{2}}\left(\xi+\frac{\Omega_r}{\omega_o^2}\right)\right]\right|_{r=r_{0}}=0.
\label{4.9}
\end{equation}
Celestial coordinates which are given by Eqs. \ref{4.8a} and \ref{4.8b} contribute to the incomplete shadow of the wormhole, as shown in the blue solid curve in Fig. \ref{fig:1}(b). The remaining part of the shadow is contributed by the unstable orbits at the throat. Therefore from Eqs. \ref{4.8a} and \ref{4.8b}, we can write,
\begin{equation}
    \alpha^2+\beta^2+\frac{\Omega_\theta}{\omega_o^2}=\xi
\label{4.10}
\end{equation}
and using expressions of $\eta$ from Eq. \ref{4.8a} and $\xi$ from Eq. \ref{4.10} into Eq. \ref{4.9}, we get,
\begin{equation}
 {\Bigg[\left(1+\alpha \omega_{T} \sin \theta\right)^{2}\frac{r^{2} K^{2}}{N^{2}}- \alpha^{2}- \frac{\Omega_r+\Omega_{\theta}}{\omega_o^2}=\left.\beta^{2}\Bigg]\right|_{r=r_{0}} }.
\label{4.11}
\end{equation}
This contributes to the shadow which is shown in the red curve in Fig. \ref{fig:1}(b), therefore the shadow will be the bounded region consisting of blue and red curves,  indicated by the solid blue and solid red curves, respectively, while disregarding the dashed red portion. Here the extreme left point of the boundary of the shadow in the celestial plane is found by setting $\beta=0$ in the expression \ref{4.11} and using $\Omega_r+\Omega_\theta=r^2K^2\omega_P^2$ from Eq. \ref{3.5}, we get,
\begin{equation}
    \alpha_L = \frac{-rK\sqrt{\frac{r^2K^2 N^2 \omega_T^2 \omega_P^2}{\omega_o^2 \sin ^2\theta}+N^2-\frac{N^4\omega_P^2}{\omega_o^2}}-r^2K^2 \omega_T\sin \theta} {r^2K^2 \omega_T^2 \sin ^2\theta-N^2}.
\label{4.12}
\end{equation}
However, this expression is not valid for the homogeneous plasma distribution which will be discussed in the next section. The wormhole shadows in vacuum are shown in Fig. \ref{fig:2} for spin parameter $a=0.99$ (left) and different inclination angles and similarly for fixed inclination angle $\theta=90^\circ$ (right) and different spin for reference purposes. 

\section{SHADOW FOR SPECIFIC PLASMA PROFILES}
In this section, our focus shifts toward exploring the effects of commonly discussed plasma distribution profiles on the shadow of rotating wormhole space-time. A crucial criterion to consider is the satisfaction of the separability condition outlined in equation \ref{3.5} while choosing the plasma distribution functions. Notably, Shapiro \cite{shapiro1974accretion} made significant advancements in accretion studies involving black holes and determined that the plasma frequency is proportional to $r^{-3/2}$ for pressureless plasma. It is imperative to acknowledge this radial decrease in plasma frequency when examining the dependence of plasma on $\theta$, especially in the case of inhomogeneous plasma distributions.

Furthermore, we must emphasize the importance of investigating a generalized form of plasma distribution. By doing so, we can highlight the distinguishing characteristics and disparities it holds when compared to other plasma distribution profiles. This comprehensive analysis enables us to gain a deeper understanding of the intricate relationship between plasma and the unique properties of rotating wormhole space-time.

\subsection{Homogeneous plasma distribution}
\begin{figure*}
\centering
\subfigure(a){\includegraphics[width = 2.25in, height=2.15in]{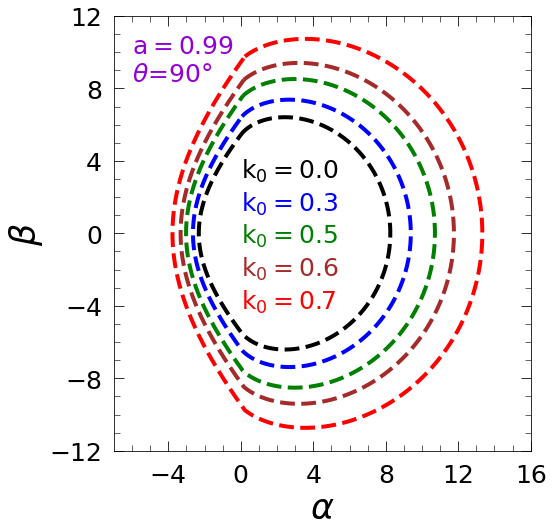}}
\subfigure(b){\includegraphics[width = 2.15in, height=2.15in]{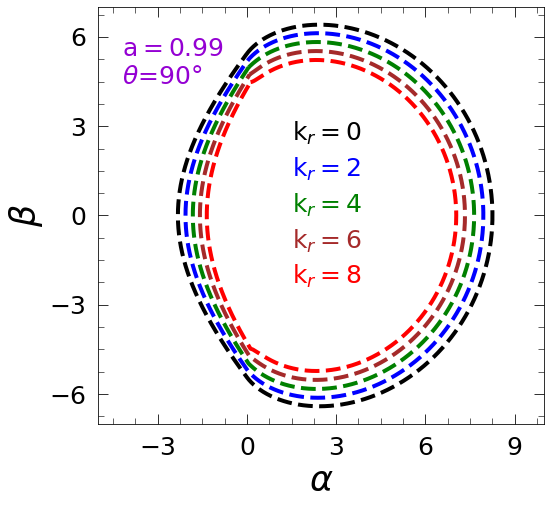}} 
\subfigure(c){\includegraphics[width = 2.15in, height=2.15in]{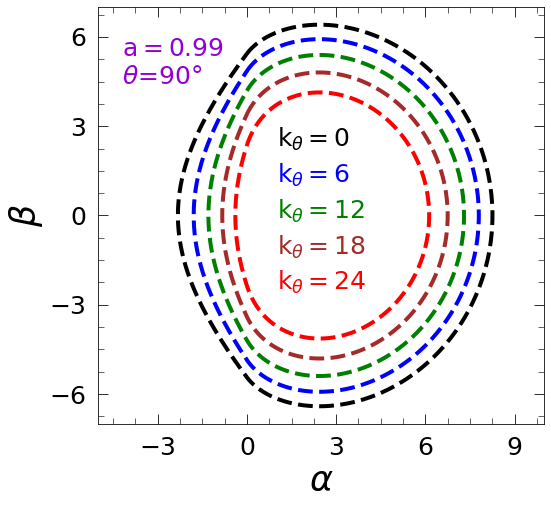}}
\subfigure(d){\includegraphics[width = 3in, height=2.15in]{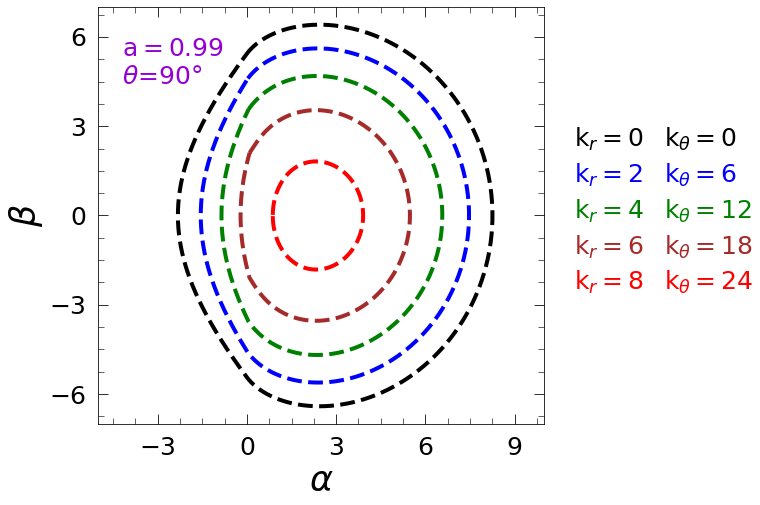}}
\caption{Comparison of Wormhole shadows with throat radius $r_0=2$, spin parameter $a=0.99$, and inclination angle $\theta=90^\circ$ with various plasma parameters ($k_0$, $k_r$, $k_\theta$) for the plasma distributions : (a) $\omega_P^2=k_0\omega_o^2$, (b) $\omega_P^2=\frac{k_r}{r^{3/2}}\omega_o^2$, (c) $\omega_P^2=\frac{k_\theta\sin^2\theta}{r^2}\omega_o^2$, and (d) $\omega_P^2=\frac{ k_r\sqrt{r}+k_\theta\sin^2\theta}{r^2}\omega_o^2$.}
\label{fig:3}
\end{figure*}

Firstly, we have considered the homogeneous plasma distribution between the observer and the source which is widely studied to understand the physical phenomena \cite{Perlick:2017fio, Bisnovatyi-Kogan_2010},
\begin{equation}
    \frac{\omega_P^2}{\omega_o^2}=k_0,
\label{5.a1}
\end{equation}
where $k_0$ denotes the homogeneous plasma parameter and it varies from $0$ to $1$ in order to satisfy the constraint Eq. \ref{2.7}. By using Eqs. \ref{3.5} and \ref{5.a1} we can write the following expressions,
\begin{equation}
    \Omega_r(r)=k_0r^2\omega_o^2 ~,~~ \Omega_{\theta}(\theta)=0.
\label{5.a2}
\end{equation}
Hence, the celestial coordinates for homogeneous plasma are given by solving Eqs. \ref{4.7a} and \ref{4.7b} as,
\begin{equation}
    \alpha = -\frac{\eta \csc \theta}{\sqrt{1-k_0}} ~,~~ \beta=\sqrt{\frac{\xi -\eta ^2 \csc ^2\theta }{1-k_0}},
\label{5.a3}
\end{equation}
and the contribution from the wormhole throat for homogeneous plasma space-time is given by 
\begin{align}
\begin{split}
 \Bigg[\left(1+\alpha \omega_T \sin \theta\sqrt{1-k_0}\right)^2\frac{r^2 K^2}{N^2(1-k_0)}- \alpha^2\\- \frac{\Omega_r}{\omega_{o}^2(1-k_0)}=\left.\beta^2\Bigg]\right|_{r=r_0} ,
 \label{5.a4}
\end{split}
\end{align}
while $\alpha_L$ as mentioned in Sec IV is found by solving $\beta=0$ , Eq. \ref{5.a3}.
In Fig. \ref{fig:3}(a), we have plotted this case for spin $a=0.99$ and it provides valuable insights into the behavior of the last photon ring, revealing that its radius expands in conjunction with larger homogeneous plasma parameters. This observation leads us to the inference that the universe is not filled with homogeneous plasma. If that were the case, we would have been able to detect these compact objects using low-resolution radio telescopes, given that the radius of the photon ring increases as the plasma parameter rises. Similar behavior has been observed for the lower spin values as well.
\subsection{Radial plasma distribution}
For this analysis, we have specifically focused on the radial plasma profile, where $\Omega_\theta$ is set to zero. We adopted the plasma profile proposed by Shapiro et al. \cite{shapiro1974accretion} to look at its effect on the rotating wormhole shadow as,
\begin{equation}
    \frac{\omega_P^2}{\omega_o^2}=\frac{k_r}{r^{3/2}},
\label{5.b1}
\end{equation}
where $k_r$ denotes the radial plasma parameter and its value should be in accordance with Eq. \ref{2.7}. We can calculate $\Omega_r$ and $\Omega_\theta$ by using Eqs. \ref{3.5} and \ref{5.b1} as,
\begin{equation}
    \Omega_r=k_r r^{1/2}\omega_o^2 ~,~ ~ \Omega_{\theta}=0,
\end{equation}
and the celestial coordinates for this plasma profile using Eqs. \ref{4.8a} and \ref{4.8b} are given by
\begin{equation}
    \alpha = -\eta  \csc\theta  ~,~~ \beta=\sqrt{\xi -\eta ^2 \csc ^2\theta }.
\end{equation}
In this scenario, the contribution of the throat to the shadow can be determined using Eq. \ref{4.11}. We have demonstrated this case in Fig. \ref{fig:3}(b) for spin parameter $a=0.99$ which illustrates the shadow of the wormhole for various radial plasma parameters, with $k_r$ equal to $0, 2, 4, 6,$ and $8$. Notably, it becomes evident that the plasma density has a negative impact on the shadow, which contrasts with the behavior observed in the case of a homogeneous plasma distribution, as depicted in Fig. \ref{fig:3}(a). As the plasma parameter increases, the shadow gradually becomes undetectable since the previously mentioned conditions (Eqs. \ref{2.7} and \ref{3.17}) are no longer satisfied. These observations shed light on the intricate relationship between plasma density and the resulting shadow characteristics.
\begin{figure*}
\subfigure{\includegraphics[width = 2.25in, height=2.15in]{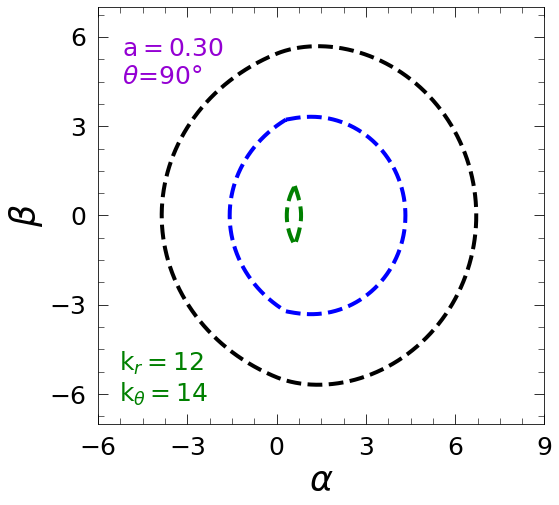}}
\subfigure{\includegraphics[width = 2.25in, height=2.15in]{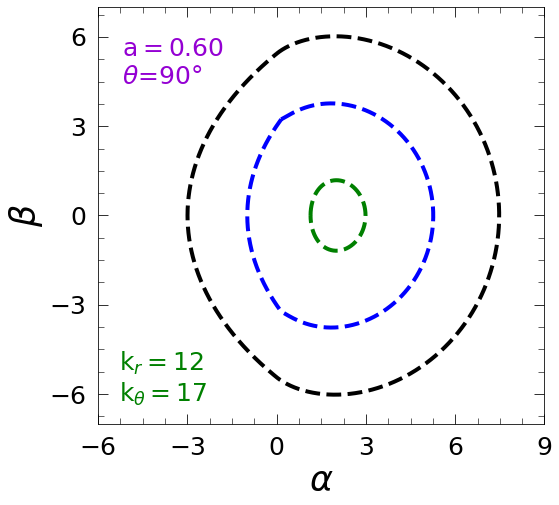}}
\subfigure{\includegraphics[width = 2.25in, height=2.15in]{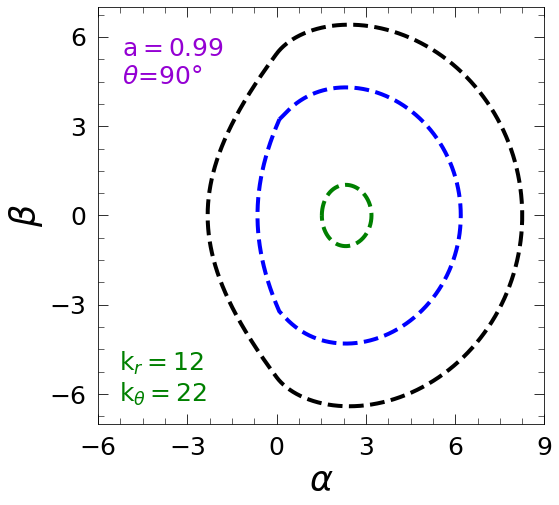}}
\subfigure{\includegraphics[width = 2.25in, height=2.15in]{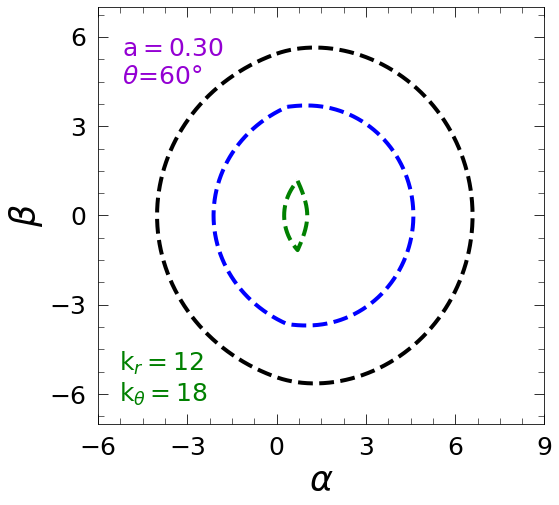}}
\subfigure{\includegraphics[width = 2.25in, height=2.15in]{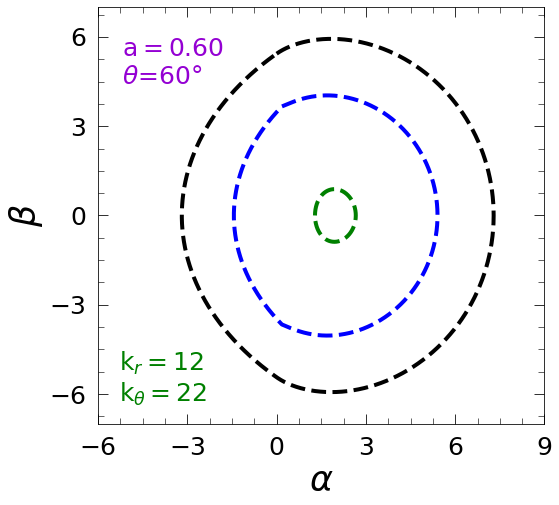}}
\subfigure{\includegraphics[width = 2.25in, height=2.15in]{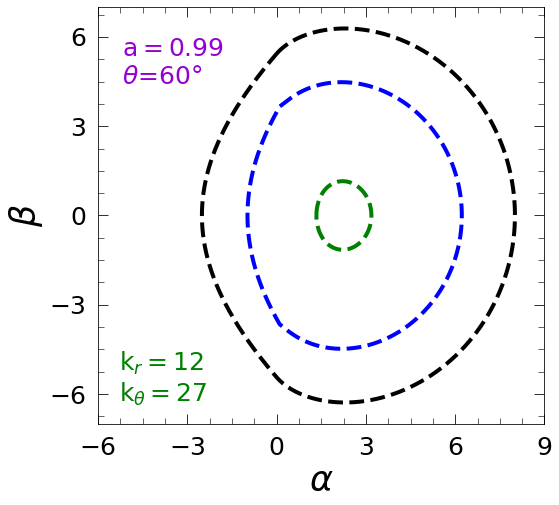}}
\subfigure{\includegraphics[width = 2.25in, height=2.15in]{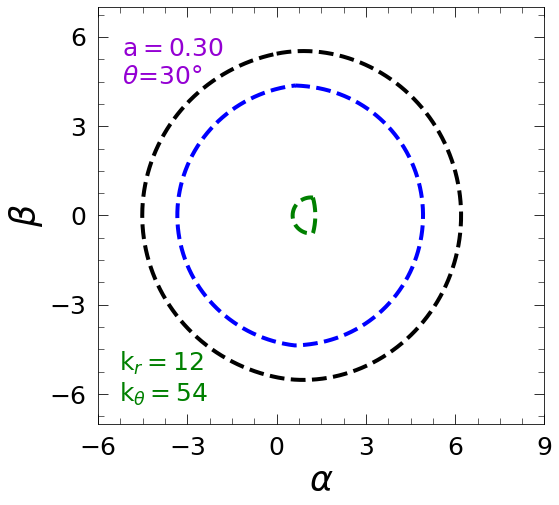}}
\subfigure{\includegraphics[width = 2.25in, height=2.15in]{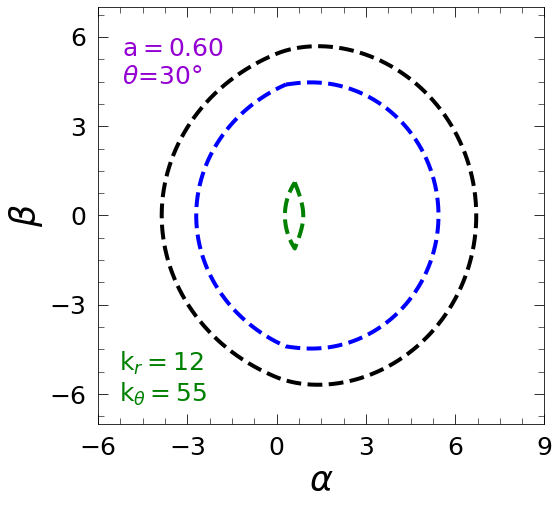}}
\subfigure{\includegraphics[width = 2.25in, height=2.15in]{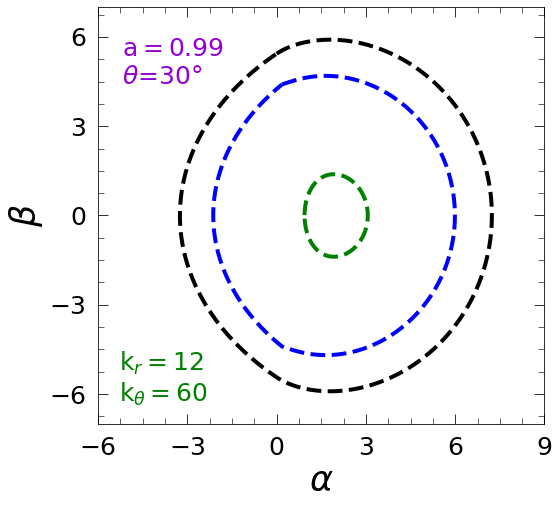}}
\caption{Wormhole shadows for generalized plasma distribution showing the disappearance of last photon ring (green curves) for various inclination angles and spin parameters with different radial plasma parameters, $k_r$ and longitudinal plasma parameter, $k_\theta$. The black curve represents the photon ring without plasma, blue curves represent shadow with low plasma parameters and are shown here for the illustration purpose of shrinking shadow size. Considering plasma profile as $\omega_P^2=\frac{ k_r\sqrt{r}+k_\theta\sin^2\theta}{r^2}\omega_o^2$.}
\label{fig:4}
\end{figure*}
\subsection{Latitudinal plasma distribution}
Now, let's explore another example where the plasma distribution is dependent on the polar ($\theta$) coordinate. In this scenario, we consider a distribution that exhibits a reduction in plasma density over increasing distances \cite{shapiro1974accretion}. This choice is essential to distinguish it from a scenario involving homogeneous plasma. To illustrate this, let's denote the plasma distribution in this case as follows:
\begin{equation}
    \frac{\omega_P^2}{\omega_o^2}=k_\theta\frac{\sin^2\theta}{r^2},
\label{5.c1}
\end{equation}
where $k_\theta$ represents the latitudinal plasma parameter, just to differentiate it from the radial plasma parameter, $k_r$ and it is chosen such that it satisfies the constraint conditions, Eq. \ref{2.7}. With the help of Eqs. \ref{3.5} and \ref{5.c1}, we can write the following expressions,
\begin{equation}
    \Omega_r=0 ~,~~ \Omega_{\theta}=k_\theta\omega_o^2 \sin^2\theta,
\end{equation}
and the celestial coordinates for this plasma profile using Eqs. \ref{4.8a} and \ref{4.8b} are given by
\begin{equation}
    \alpha = -\eta  \csc\theta  ~, ~~  \beta=\sqrt{\xi -\eta ^2 \csc ^2\theta -k_\theta\sin^2\theta}.
\end{equation}
Fig. \ref{fig:3}(c) provides a comparative visualization of the shadow cast by the wormhole for this particular plasma profile. Remarkably, it becomes evident that the dependence of plasma density on the $\theta$ coordinate exerts a significant influence on the size of the shadow, surpassing the impact of the radial profile. This finding is particularly noteworthy, as previous studies primarily concentrated on radial profiles \cite{abdujabbarov2016shadow} and omitted the analysis of such latitudinal profiles. It underscores the importance of considering a generalized plasma density distribution to gain a more profound understanding of the shadow boundary in plasma space-time. Therefore, by incorporating the influence of plasma density variation with respect to $\theta$, we can delve deeper into the intricacies of shadow formation and unravel more comprehensive insights into the behavior of wormholes in the presence of varying plasma distributions. Motivated by this, in the next subsection, we will be studying the more general case for plasma distribution.
\subsection{Generalized plasma distribution}
Now, let's consider the more comprehensive scenario where the plasma distribution depends on both the radial coordinate ($r$) and the angular coordinate ($\theta$). This broader analysis allows us to gain further insights into the effects of plasma densities on wormhole shadows. For this case, we denote the plasma distribution as follows,
\begin{equation}
    \frac{\omega_P^2}{\omega_o^2}=\frac{ k_r\sqrt{r}+k_\theta\sin^2\theta}{r^2}.
\label{5.d1}
\end{equation}
This particular profile is essentially a combination of the two previously discussed profiles. It incorporates the additive contributions from each of them. We adopt a similar plasma distribution profile to the one proposed by Perlick et al. \cite{Perlick:2017fio}, which provides valuable insights into the behavior of the plasma distribution in relation to the formation of wormhole shadows. Now, using Eqs. \ref{3.5} and \ref{5.d1},
\begin{equation}
    \Omega_r=k_rr\omega_o^2 ~, ~~ \Omega_{\theta}=k_\theta \omega_o^2 \sin^2\theta,
\end{equation}
and the celestial coordinates for this generalized plasma profile using Eqs. \ref{4.8a} and \ref{4.8b} are given by
\begin{equation}
    \alpha = -\eta  \csc\theta  ~,~ ~  \beta=\sqrt{\xi -\eta ^2 \csc ^2\theta -k_\theta\sin^2\theta}.
\end{equation}
Fig. \ref{fig:3}(d) showcases the shadow cast by the wormhole for different values of $k_r$ and $k_\theta$. Interestingly, the generalized plasma profile demonstrates superior performance compared to the other two profiles previously discussed. This emphasizes the significance of studying the more comprehensive generalized plasma profile rather than solely focusing on the radial profile. A similar kind of behavior has been observed for the low spin values. Hence, we have chosen to compare the different plasma density profiles with the higher spin case.
Furthermore, it is noteworthy that the individual shadows resulting from the specific plasma parameters are larger when compared to their combined effect. As we delve deeper into the analysis, we observe that the shadow progressively diminishes in size with increasing plasma parameters. At a certain critical value, the shadow may eventually vanish or become undetectable altogether. This phenomenon has been shown in Fig. \ref{fig:4} for different values of spin parameters as well as at different inclination angles. It can be observed that as the plasma parameter increases, the wormhole shadow started shrinking and eventually disappears which has been shown by dashed green curves. Please note that the plasma parameter values corresponding to these green curves do not serve for maximum value after which the shadow gets disappear completely.

By considering these generalized plasma distributions and closely examining the changes in the resulting shadows, we can potentially gain valuable insights into the plasma distribution along the observational path. 
This motivates us to further look into the photon trajectories in plasma space-time. Therefore, in the next section, we will be exploring the weak gravitational lensing within the effect of the plasma distribution around the rotating wormhole space-time.

\section{Weak gravitational lensing}

In this section, we explore the influence of plasma distributions on the deflection angle within the framework of the weak field approximation. As we know, when light rays traverse the vicinity of massive objects, they experience deviations from their original paths. Here, we present the analytical expression for the deflection angle, focusing specifically on the case when the observer is situated in the equatorial plane ($\theta=90^\circ$) of the source.
\begin{figure*}
\subfigure(a){\includegraphics[width = 2.9in, height=2.7in]{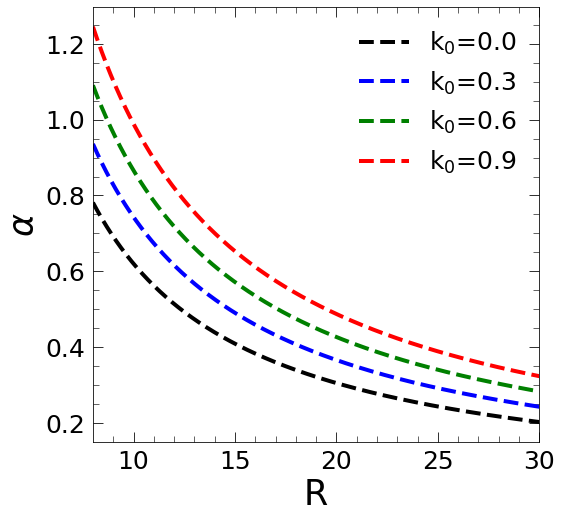}}
\subfigure(b){\includegraphics[width = 2.9in, height=2.7in]{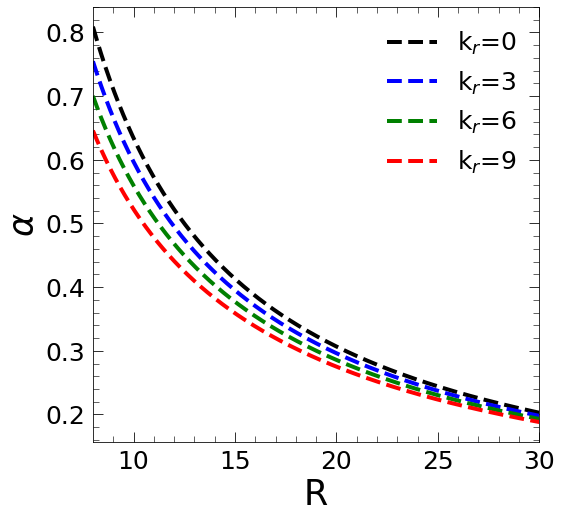}}
\caption{Weak deflection angle as a function of closest distance (R) with spin parameter, $a=0.5$ in Teo wormhole space-time for plasma profiles: (a) $\omega_P^2=k_0\omega_o^2$, and (b) $\omega_P^2=\frac{k_r}{r^2}\omega_o^2$.}
\label{fig:5}
\end{figure*}
\begin{table*}[ht]
\addtolength{\tabcolsep}{25pt}
\fontsize{13.5pt}{10.25pt}
\centering
\renewcommand{\arraystretch}{2.5}
\caption{Deflection angle in homogeneous and non-homogeneous plasma distributions for Teo wormhole space-time.}
\begin{tabular}{|c|c|}
\hline
\text{\fontsize{12}{14}\selectfont Plasma distribution} & \text{\fontsize{12}{14}\selectfont Deflection angle} \\
\hline
 $\frac{\omega_P^2}{\omega_o^2}=k_0$ & $\pm \left(\frac{3r_0}{R}+\frac{4a}{R^2}+\left(\frac{2r_0 k_0}{R}+\frac{2a k_0}{R^2}\right)\right)$ \\

 $\frac{\omega_P^2}{\omega_o^2}=\frac{k_r}{r^2}$ & $\pm \left(\frac{3r_0}{R}+\frac{4a}{R^2}+\frac{ r_0 a (9\pi-14)}{R^3}+\frac{k_r}{2R^2}\left(\frac{r_0 (2\pi-3)} {R}-\pi \right)\right)$ \\
\hline
\end{tabular}

\label{Table 1}
\end{table*}
By examining the effects of plasma distributions on the deflection angle, we can gain a deeper understanding of how the presence of plasma affects the trajectory of light rays near massive objects. This analysis allows us to investigate the intricate interplay between plasma and gravity, shedding light on the nature of weak gravitational lensing in the presence of plasma. The derived analytic expression provides a valuable tool for predicting and analyzing the deflection of light in various astrophysical scenarios, contributing to our overall comprehension of the behavior of light in the presence of massive objects and plasma distributions. In order to analyze the deflection angle, we first calculate the geodesic equations with the help of Hamilton's equation \ref{2.6} and the Hamiltonian for rotating plasma space-time given by Eq. \ref{2.1} as,
\begin{equation}
    \dot{\phi}=\frac{\partial H}{\partial p_\phi}=g^{t \phi} p_t+g^{\phi \phi} p_\phi, ~,~~~ \dot{r}=\frac{\partial H}{\partial p_r}=g^{r r} p_r,
\label{6.1}
\end{equation}
which can be further simplified as follows,
\begin{equation}
    \left(\frac{\dot{r}}{\dot{\phi}}\right)^2=\left(\frac{g^{r r} p_r}{g^{\phi \phi} p_\phi+g^{\phi t} p_t}\right)^2.
\label{6.2}
\end{equation}
Now, for the massless particles, the Hamiltonian should be zero ($H=0$). Therefore, Eq. \ref{2.1} can be written by using the definitions such that $p_t=-\omega_o, ~\text{and}~ p_\theta=L$ as,
\begin{equation}
      g^{rr} p_r^2=-\left(g^{t t} \omega_o^2-2 g^{\phi t} \omega_o L+g^{\phi \phi} L^2+\omega_P^2\right).
\label{6.3}
\end{equation}
Hence Eqs. \ref{6.2} and \ref{6.3} can be simplified to,
\begin{align}
\begin{split}
    \left(\frac{\dot{r}}{\dot{\phi}}\right)^2=-\frac{g^{r r}}{\left(g^{\phi \phi} L-g ^{\phi t}\omega_o\right)^2}(g^{t t} \omega_o^2-2 g ^{\phi t}\omega_o L\\+g^{\phi \phi} L^2+\omega_P^2),
    \end{split}
\label{6.4}
\end{align}
and to simplify the above Eq. we have considered the following definitions,
\begin{equation}
    \frac{\omega_o}{L}=\lambda ~,\quad \frac{\omega_P^2}{\omega_o^2}=X,
\label{6.5}
\end{equation}
thus, equation \ref{6.4} can be modified as,
\begin{equation}
   \left(\frac{\dot{r}}{\dot{\phi}}\right)^2=\frac{-g^{r r}}{\left(g^{\phi \phi}-g^{\phi t} \lambda\right)^2}\left(g^{t t} \lambda^2-2 g^{\phi t} \lambda+g^{\phi \phi}+ X \lambda^2\right).
\label{6.6}
\end{equation}
Since the deflection angle is calculated when the light deviates from its original path and consequently when it is at the closest approach ($r=R$) to the central object. Therefore, at the closet distance, we can define,
\begin{equation}
    \left.\left(\frac{\dot{r}}{\dot{\phi}}\right)\right|_{r=R}=0.
\label{6.7}
\end{equation}
Now, the evaluation needs to be done at $r=R$ as mention in Eq. \ref{6.7}, we have considered the following expressions for the metric and plasma functions, being evaluated at $r=R$:
\begin{equation}
\begin{split}
    \left.g^{t t}\right|_R=G^{t t},\left.g^{\phi t}\right|_R=G^{\phi t},\left.g^{\phi \phi}\right|_R=G^{\phi \phi},\\ \left.g^{r r}\right|_R=G^{rr}, \left.X\right|_R=Y.
\end{split}
\label{6.8}
\end{equation}
\begin{figure*}
\subfigure(a){\includegraphics[width = 2.9in, height=2.7in]{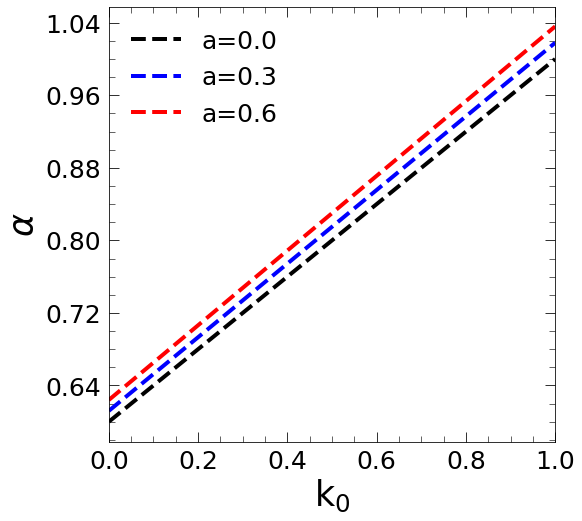}}
\subfigure(b){\includegraphics[width = 2.9in, height=2.7in]{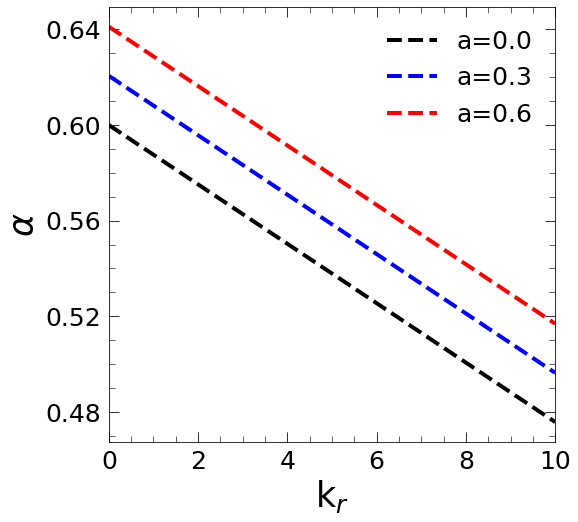}}
\caption{Weak deflection angle in Teo wormhole space-time for closest distance approach, $R=10$ as a function of plasma parameters ($k_0,~k_r$) for various spin parameters in case of plasma profiles: (a) $\omega_P^2=k_0\omega_o^2$, and (b) $\omega_P^2=\frac{k_r}{r^2}\omega_o^2$.}
\label{fig:6}
\end{figure*}
Thus, the impact parameter, $\lambda$ is calculated by using the expressions given in Eq. \ref{6.8} with the help of Eqs. \ref{6.6} and \ref{6.7}, and given as,
\begin{equation}
\lambda=\frac{2 G ^{\phi t} \pm \sqrt{(2 G ^{\phi t})^2-4 G ^{\phi \phi}\left(G^{t t}+Y\right)}}{2\left(G^{t t}+Y\right)},
\label{6.9}
\end{equation}
and finally, the integral form for the deflection angle of the light from its original trajectory can be given by solving further using Eqs. \ref{6.6} and \ref{6.9} as,
\begin{widetext}
\begin{equation}
 \int_0^{\bar{\alpha}} d \phi=\pm 2 \int_{-\infty}^{\infty}\left[\frac{-g^{r r}}{\left(g^{\phi \phi}-g^{\phi t} \lambda\right)^2}\left((g^{t t}+X) \lambda^2-2 g^{\phi t} \lambda+g^{\phi \phi}\right)\right]^{-1 / 2} d r.
 \label{6.10}
\end{equation}
\end{widetext}


It is important to note that the deflection angle for the light following its original trajectory will be $\pi$ given that the center of coordinates corresponds to the compact object. Therefore, the actual deflection angle is determined by $\alpha=\bar{\alpha}-\pi$.

In the subsequent analysis, we proceed to calculate the deflection angles for both homogeneous and non-homogeneous plasma distributions. The homogeneous plasma distribution is characterized by uniform plasma density, while the radial plasma distribution exhibits a density variation in the non-homogeneous direction. By studying these specific cases, we can discern the effects of plasma distributions on the deflection of light and deepen our understanding of gravitational lensing phenomena in the presence of plasma. We have considered the following plasma distributions \cite{Bisnovatyi-Kogan_2010}.
\begin{equation}
    \frac{\omega_P^2}{\omega_o^2}=k_0~,~~\frac{\omega_P^2}{\omega_o^2}=\frac{k_r}{r^2}.
\label{6.11}
\end{equation}
In the case of a homogeneous plasma distribution, the values of $k_0$ fall within the range of ($0,1$) as discussed in section V(A). Additionally, the choice of $k_r$ is determined to satisfy Eq. \ref{2.7} which takes into account the gravitational redshift. It is worth noting that in previous studies \cite{bisnovatyikogan2015gravitational, farruh2021}, the redshift condition has often been neglected. However, it is crucial to consider this condition as it significantly influences the trajectory of light and, consequently, the deflection angle.

We have considered the weak field limit and lower plasma densities for simplicity to calculate the weak deflection angle for the slow-rotating wormhole. The detailed derivation has been performed in Appendix \ref{Appendix A}. The resulting values for the deflection angle are presented in Table \ref{Table 1}, providing a comprehensive overview of the deflection angles for both the homogeneous and non-homogeneous plasma profiles. By examining the values of the deflection angle, we can gain insights into the effects of plasma distributions on the path of light in the vicinity of massive objects. Therefore, we examine both the homogeneous and non-homogeneous plasma distributions around the rotating wormhole geometry and studied their effect on the deflection angle of the light rays.

 

The deflection angle exhibits a decrease with the closest distance to the wormhole (See Fig. \ref{fig:5}(a) and  \ref{fig:5}(b)), indicating a reduced gravitational influence. It is noteworthy that at higher plasma densities, the deflection angle increases, as illustrated in Fig. \ref{fig:6}(a) for all values of the spin parameter. This observation gives validation of the earlier observed phenomenon such that the shadow radius increases with the plasma density in uniform plasma distribution (see Fig. \ref{fig:3}(a)).

In the case of a non-homogeneous plasma distribution, an intriguing observation is that the deflection angle decreases with increasing plasma densities, as shown in Figs. \ref{fig:5}(b) and \ref{fig:6}(b) for all values of the spin parameter. This stands in contrast to the homogeneous case and provides an explanation for the negative impact of plasma on the shadow which already has been observed in the case of the shadow (see Fig. \ref{fig:3}(b)). Notably, the influence of non-homogeneous plasma distributions on the deflection angle has not been extensively explored in previous studies. Most investigations on the effects of plasma on the deflection angle by compact objects have focused on a single isothermal sphere model, commonly employed for galaxy modelling which yielded a positive impact of plasma on the deflection angle \cite{farruh2021}.

It is important to note that the choice of the plasma parameter value should ensure low plasma density and compliance with the condition given by Eq. \ref{2.7}. In previous studies, researchers have typically considered plasma parameter values ranging from 0 to 1 \cite{farruh2021}. However, within the given impact parameter constraints along with the condition given by Eq. \ref{2.7}, a range of plasma parameter values can be chosen to study the deflection angle. Therefore, by analyzing the effects of plasma on the deflection angle, we can gain valuable insights into the distribution of plasma in the vicinity of compact objects. This investigation serves as a powerful tool for studying and understanding the properties of plasma surrounding these intriguing cosmic structures.

\section{CONSTRAINING THE WORMHOLE SHADOW AND PLASMA PARAMETERS}

To determine the plasma parameters and size of the wormhole, we employ observational data released by EHT \cite{EHT2019_1} for the supermassive black hole located at the center of the elliptical galaxy Messier 87 (M87), also known as M$87^{\ast}$. By examining the average angular size of the shadow and its deviation from circularity, we can constrain shadow and plasma parameters. As the shadow possesses reflection symmetry around the $\alpha$-axis in the celestial plane, we calculate its geometric center $(\alpha_0, \beta_0)$ using the integrals $\alpha_0=1 / A \int \alpha d A$ and $\beta_0=0$. Here, $dA$ represents an area element. Next, we introduced an angle $\phi$ defined as the angle between the $\alpha$-axis and the vector connecting the geometric center $(\alpha_c, \beta_c)$ with a point $(\alpha, \beta)$ on the boundary of the shadow. This angle $\phi$ provides valuable information for our analysis as can be seen as follows. Therefore, the average radius ($R$) of the shadow is given by \cite{Rahaman_2021},
\begin{equation}
    R^2=\frac{1}{2 \pi} \int_0^{2 \pi} l^2(\phi) d \phi,
\label{7.1}
\end{equation}
\begin{figure*}
\subfigure(a){\includegraphics[width = 3.3in, height=2.75in]{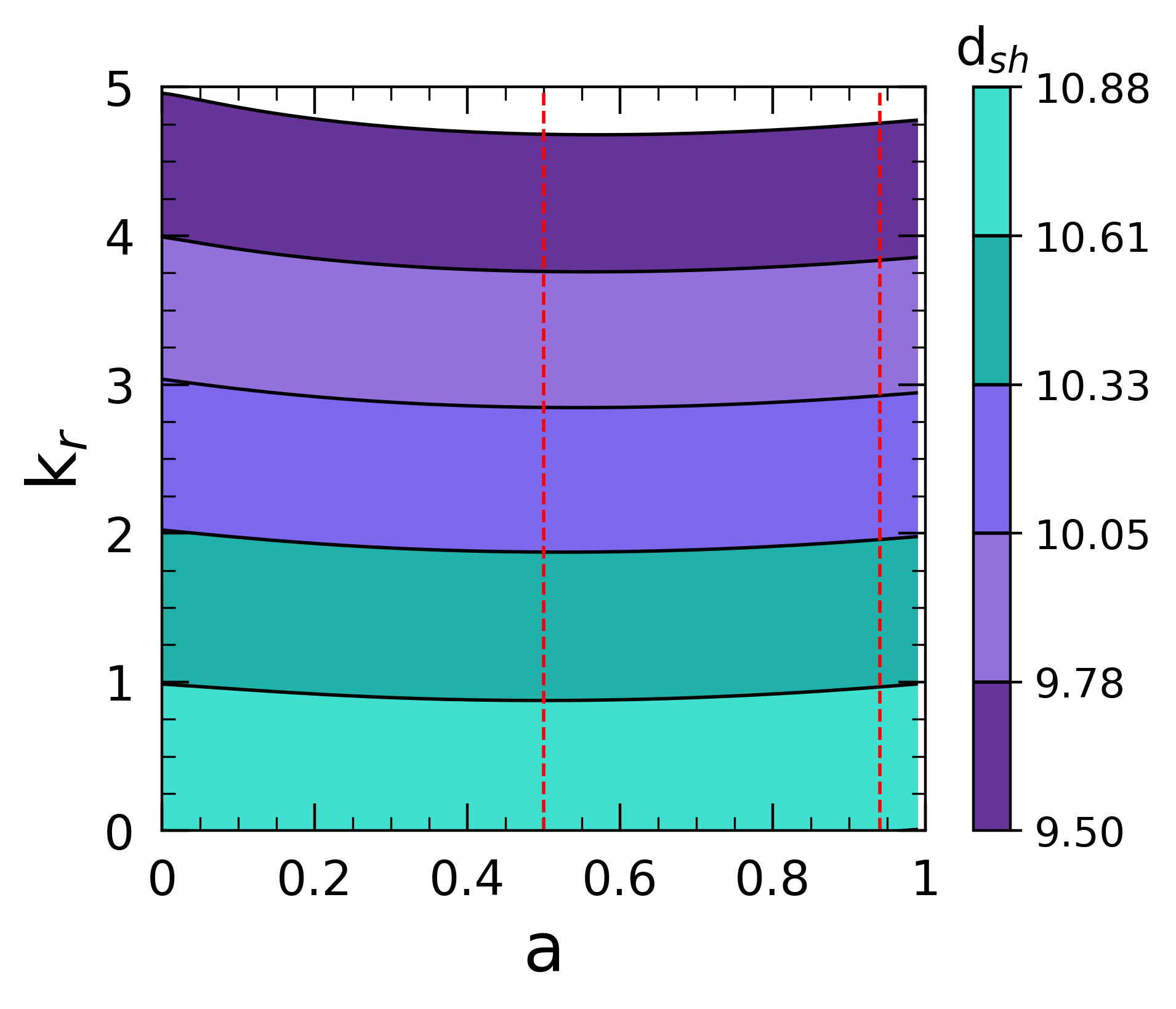}}
\subfigure(b){\includegraphics[width = 3.3in, height=2.75in]{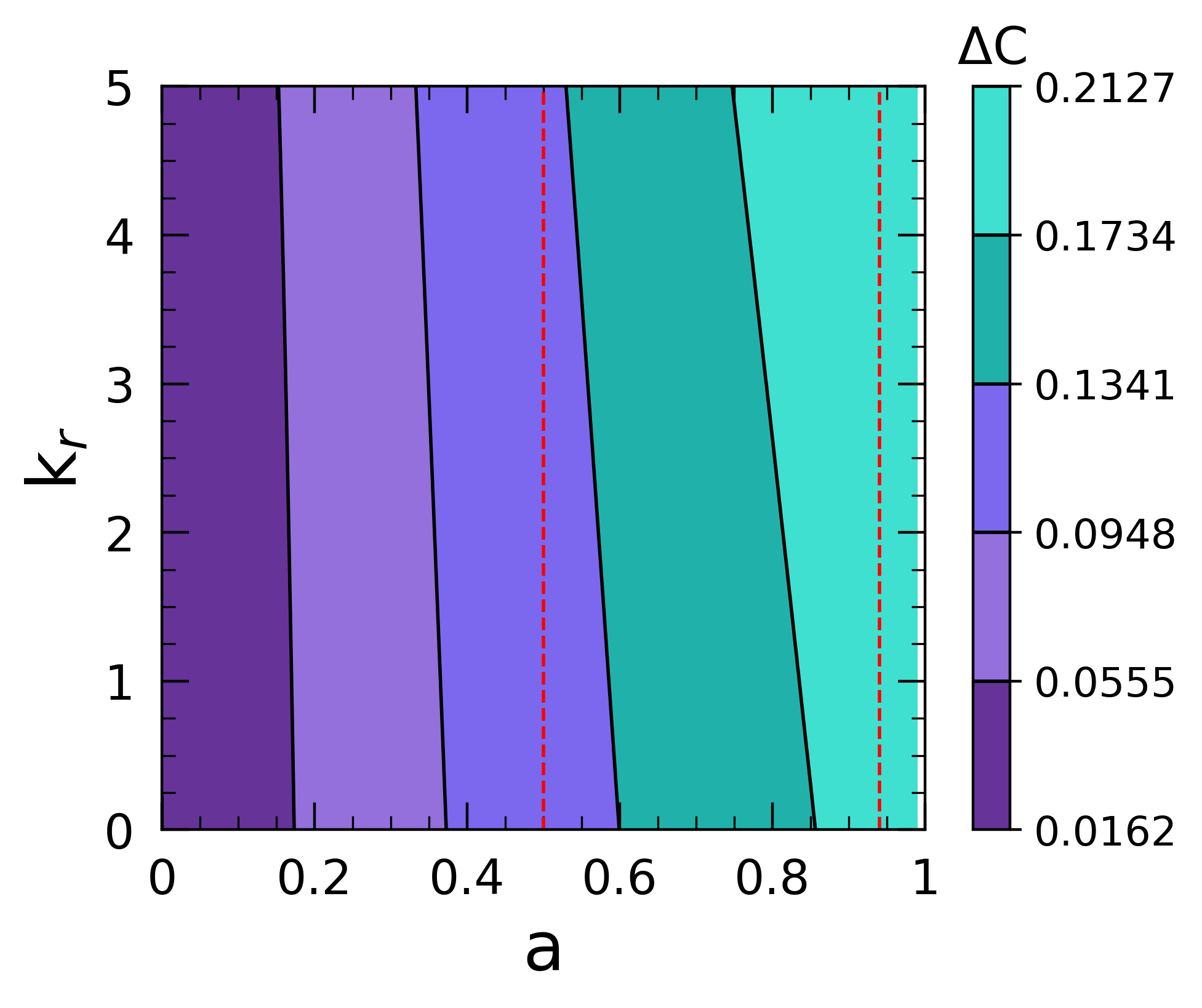}}
\caption{ Dependence of (a) the angular size and (b) the deviation of the shadow on radial plasma parameter ($k_r$) and spin with $\theta=17^\circ$ and $r_0=2$. The two red dashed lines indicate the spin range $0.5\leq a \leq 0.94$. Considering plasma profile as $\omega_P^2=\frac{k_r}{r^{3/2}}\omega_o^2$.}
\label{fig:8}
\end{figure*}
\begin{figure*}
\subfigure(a){\includegraphics[width = 3.3in, height=2.75in]{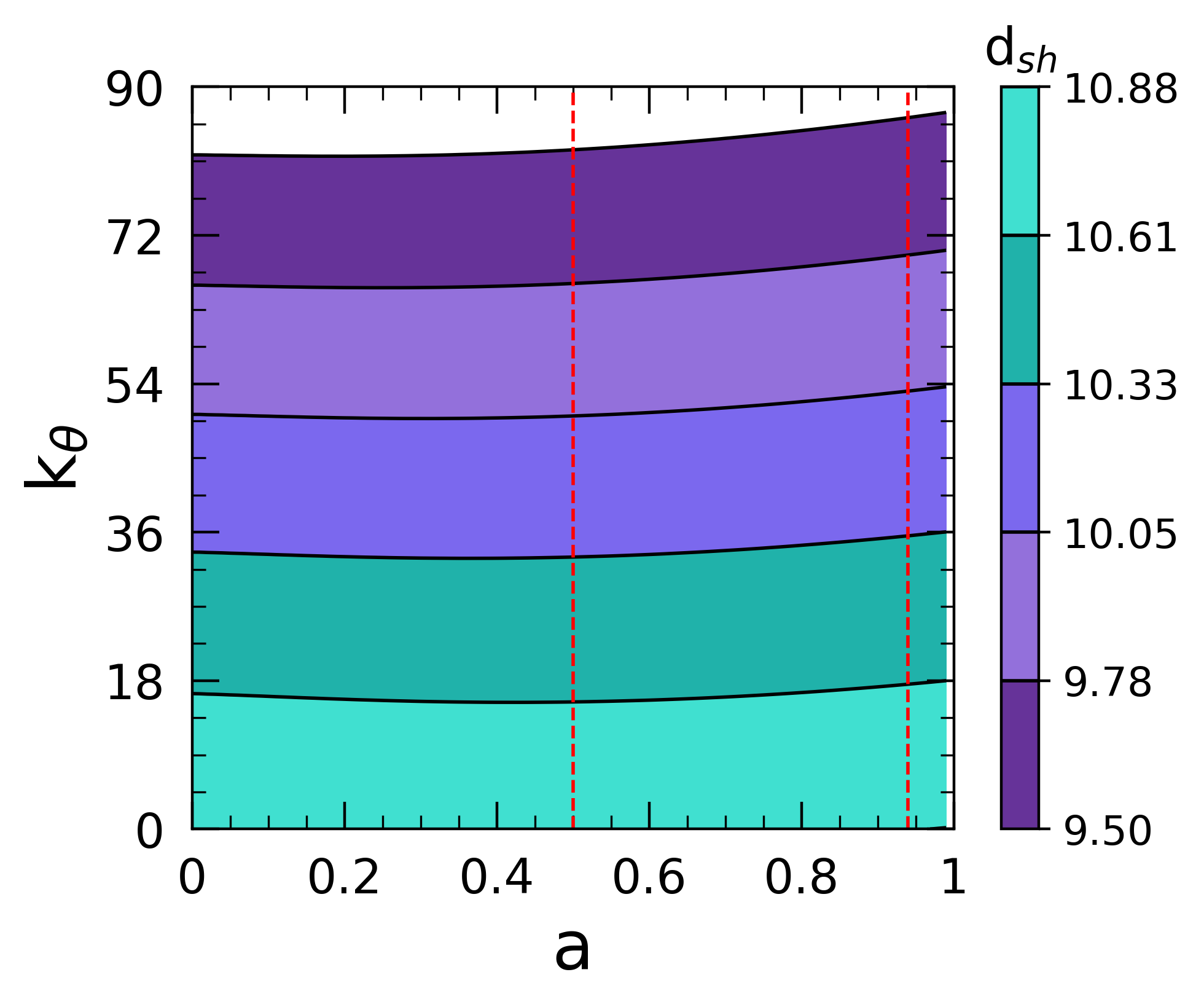}}
\subfigure(b){\includegraphics[width = 3.3in, height=2.75in]{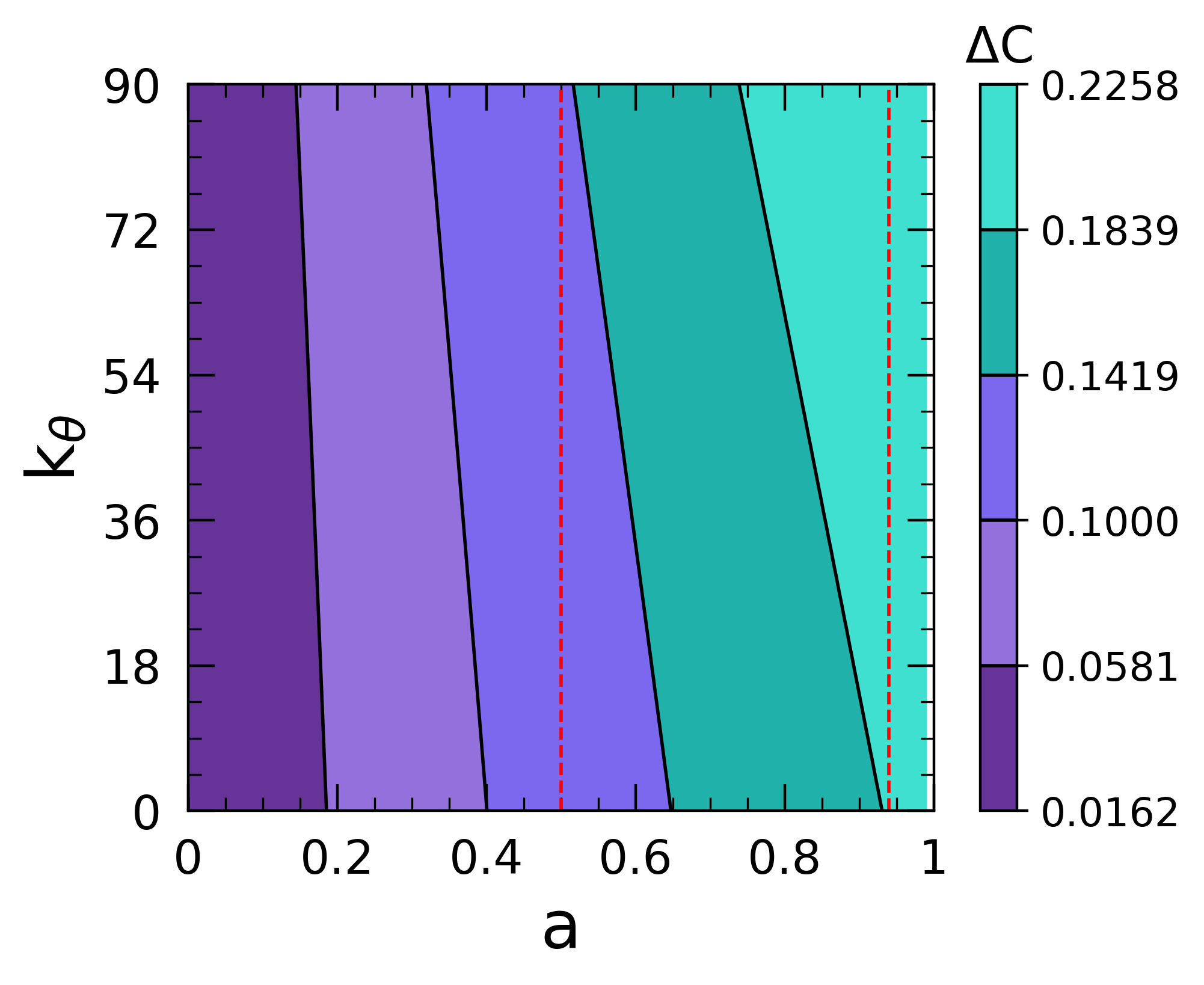}}
\caption{Dependence of (a) the angular size and (b) the deviation of the shadow on latitudinal plasma parameter ($k_\theta$) and spin with $\theta=17^\circ$ and $r_0=2$. The two red dashed lines indicate the spin range $0.5 \leq a \leq 0.94$. Considering plasma profile as $\omega_P^2=\frac{k_\theta\sin^2\theta}{r^2}\omega_o^2$.}
\label{fig:9}
\end{figure*}
where $l(\phi)=\sqrt{\left(\alpha(\phi)-\alpha_0\right)^2+\beta(\phi)^2}$ and $\phi=\tan ^{-1}\left(\beta(\phi) /\left(\alpha(\phi)-\alpha_0\right)\right)$. Following \cite{EHT2019_1}, we define the deviation $\Delta C$ from circularity as \cite{Rahaman_2021},
\begin{equation}
    \Delta C=\frac{1}{R} \sqrt{\frac{1}{2 \pi} \int_0^{2 \pi}\left(l(\phi)-R\right)^2 d \phi}.
\label{7.2}
\end{equation}
We should note that $\Delta C$ represents the fractional root mean square distance from the average radius of the observed shadow. Based on the findings of the EHT collaboration \cite{EHT2019_1}, the angular size of the observed shadow is determined to be $\Delta \theta_{\text{sh}} = 42 \pm 3 \mu$as, with a deviation $\Delta C$ of less than $10\%$. Additionally, following the same study \cite{EHT2019_1}, we adopt the distance to M87* as $D = (16.8 \pm 0.8)$ Mpc and the mass of the object as $M = (6.5 \pm 0.7) \times 10^9 M_{\odot}$. With these values, we can estimate the average diameter of the shadow by considering the angular size,
\begin{equation}
\text{Diameter} = d_{sh}=2 \times D \times \tan\left(\frac{\Delta \theta_{\text{sh}}}{2}\right)=11.0\pm1.5.
\label{7.3}
\end{equation}
This uncertainty in the shadow diameter can be determined by carefully propagating the uncertainties associated with both the distance and angular size measurements. By accounting for these uncertainties, we can obtain more reliable estimates of the shadow's diameter, enabling us to extract valuable information about the physical characteristics of M$87^{\ast}$ and delve into the intricacies of the shadow phenomenon. Therefore, these insights pave the way for further investigations and contribute to our ongoing exploration of the enigmatic nature of M$87^{\ast}$ and its surrounding environment.

The determination of the average diameter and the deviation from the circularity of the observed shadow involves considering the errors in a combined manner, with the uncertainties added in quadrature. It is important to ensure that this calculated quantity matches the expected value of the diameter (2R). Fig. \ref{fig:8} presents the results depicting the average diameter (left) and circularity deviation (right) of the shadow, considering various values for the spin parameter and the size of the wormhole throat concerning the radial plasma parameter ($k_r$). In our analysis, we have taken an inclination angle of $\theta=17^{\circ}$, which represents the angle between the jet axis and the line of sight to M$87^{\ast}$. 
\begin{figure*}
\subfigure(a){\includegraphics[width = 3.3in, height=2.75in]{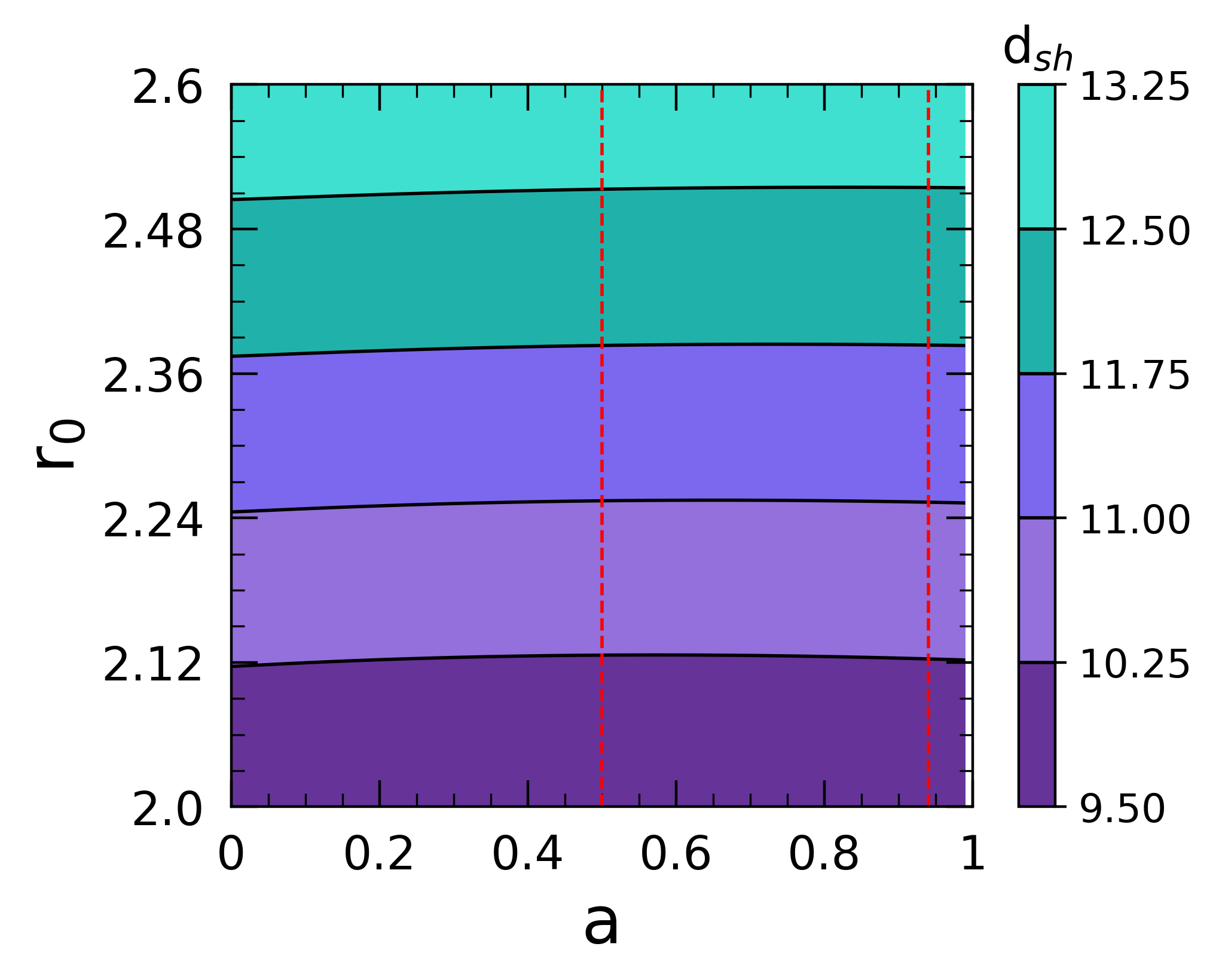}}
\subfigure(b){\includegraphics[width = 3.3in, height=2.75in]{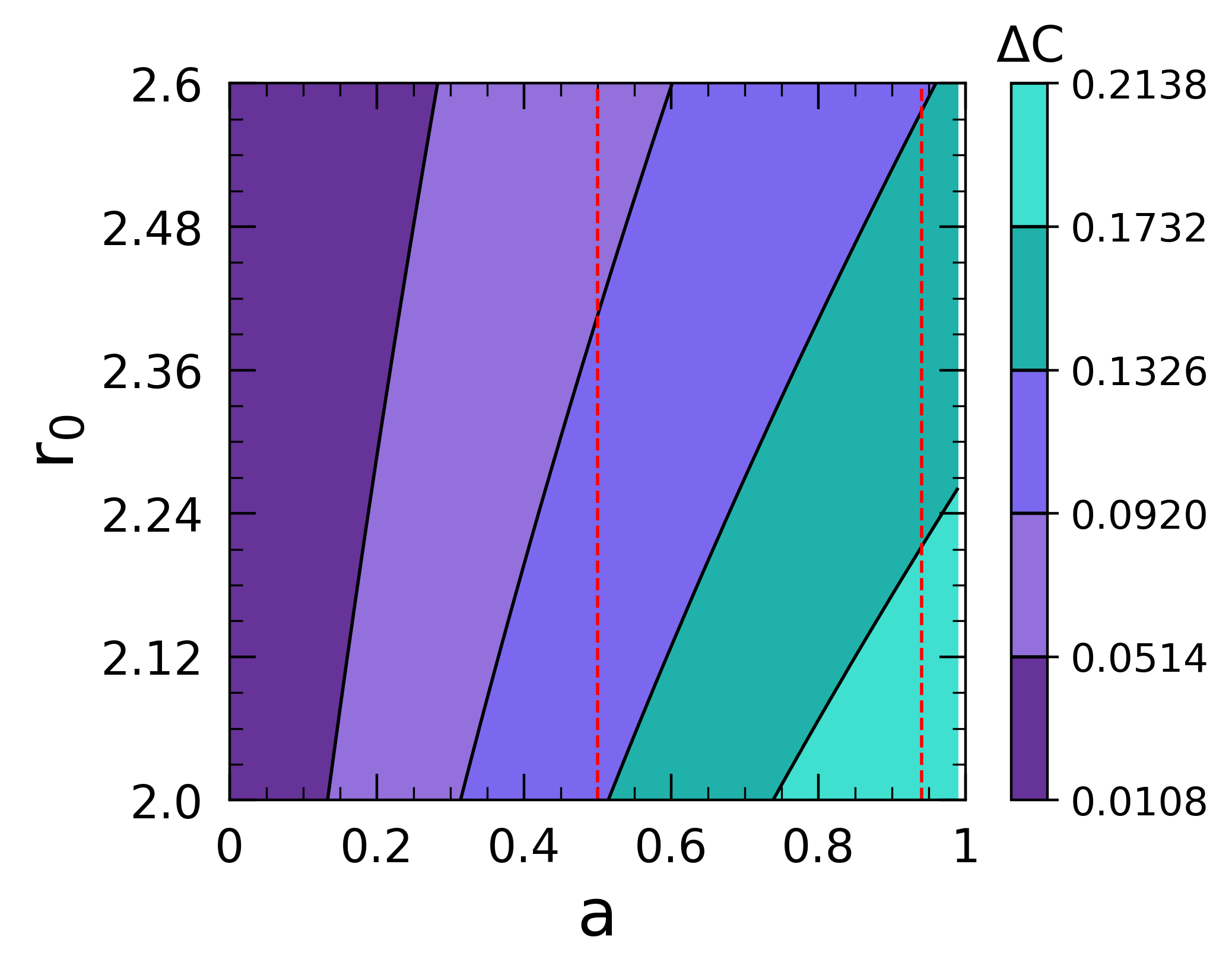}}
\caption{ Dependence of (a) the angular size and (b) the deviation of the shadow on throat size and spin with $\theta=17^\circ$, $k_r=3.5$ and $k_\theta=20.4$. The two red dashed lines indicate the spin range $0.5\leq a \leq 0.94$. Considering plasma profile as $\omega_P^2=\frac{ k_r\sqrt{r}+k_\theta\sin^2\theta}{r^2}\omega_o^2$.}
\label{fig:10}
\end{figure*}
Additionally, based on the findings of the EHT collaboration, the spin parameter falls within the range of $0.5 \leq a \leq 0.94$ which has been shown in the Fig. \ref{fig:8} with vertical red lines.

Now, to determine the maximum plasma parameters that apply to M$87^{\ast}$, we analyzed the plasma density corresponding to the smallest observed shadow size. Initially, by looking only at the radial plasma profile, we observed that the maximum value for the radial plasma parameter is $k_{rc}=4.75$ when the shadow diameter matches with the observed M$87^{\ast}$ diameter (Fig. \ref{fig:8}). Similarly, the contour line corresponding to the lowest shadow diameter Fig. (\ref{fig:9}) indicates that the maximum longitudinal plasma parameter is $k_{\theta c}=85.5$. furthermore, to cover the complete range of shadow diameters between $9.5$ and $12.5$ with the uncertainty in the M$87^{\ast}$ shadow diameter, we considered $k_r=3.5$ and $k_\theta=20.4$ in Fig. \ref{fig:10}. These choices allow us to place constraints on the maximum possible value of the throat size ($r_{0c}=2.51$) which corresponds to the maximum shadow diameter as shown in Fig. \ref{fig:10}(a). It is worth noting that the deviation from circularity ($\Delta C$) in the wormhole shadows provides interesting features about the plasma parameters. From Figs. \ref{fig:8}(b) and \ref{fig:9}(b), we observed that $\Delta C \leq 10\%$ required the spin range for M$87^{\ast}$ to be less than $0.5$ when considering either the radial or latitudinal plasma profiles. However, in the case of the generalized plasma distribution the observed circularity deviation from the shadow of M$87^{\ast}$ ($\Delta C \leq 10\%$) can still be achieved for the known spin range while obtaining shadow diameters within the observed range for M$87^{\ast}$. It is important to note that this analysis restricts the throat size such that it should not exceed the value of $r_{0c}$ to maintain $\Delta C < 10\%$ and maximum shadow diameter of 12.5 within the allowed range of spin, the value of $r_{0c}$ should be less than 2.51 for the chosen value of $k_r$ and $k_\theta$ in case of generalized plasma distributions (See Fig. \ref{fig:10}(a) and \ref{fig:10}(b)). Thus, the constraint on the maximum value of throat size may vary depending on the choice of plasma parameters. These findings further contribute to our understanding of the possible spin range and plasma distributions associated with the observed shadow of M$87^{\ast}$. Please note that we have only considered those plasma profiles which have been discussed in Section V.

Therefore, this study ascertains the range of plasma parameters and throat size that are consistent with the observed shadows and provides further insights into the properties of the M$87^{\ast}$ system.

\section{Conclusion}
In this study, our focus was on investigating the behavior of null geodesics in non-magnetized pressureless plasma within the context of a rotating wormhole space-time. We specifically examined the gravitational influence of the wormhole while neglecting the direct gravitational effects of the plasma. Instead, we considered only the dispersive properties of the plasma, affecting the trajectory of light rays.

One key finding of our work, as discussed in Section III, is the requirement of a specific plasma distribution profile to establish a generalized Carter's constant. We also emphasized the importance of including potential contributions from both inside and outside the wormhole throat, as elaborated in Section IV. Furthermore, we derived analytical formulas for the boundary of the shadow for various plasma profiles in Section V. Notably, our results revealed that the shadow size decreases with increasing plasma density. Eventually, for certain upper limits of the plasma parameters, the shadow completely disappears.

Our primary objective throughout this study was to obtain an analytical expression for the shadows observed in plasma space-time. By investigating the behavior of light rays in the presence of plasma, we aimed to enhance our understanding of the intricate interplay between gravitational and plasma effects in astrophysical phenomena. In Section VI, we conducted calculations to determine the deflection angle on a rotating wormhole in plasma space-time. Gravitational lensing phenomena have significant implications for astrophysical observations, and our study shed light on the impact of plasma on the deflection angle. Interestingly, we observed that as the plasma parameter increases, the deflection angle decreases in a non-homogeneous plasma space-time, contrary to the behavior observed in a homogeneous plasma profile. This intriguing result underscores the importance of further investigating the observational aspects and exploring the plasma distribution near compact objects.

Finally, we proceeded to constrain the size of the throat and plasma parameters mentioned in Section V by utilizing the observational data coming from EHT for M$87^{\ast}$. Our analysis revealed that the maximum allowed throat radius is determined to be $r_{0c}=2.51$, which corresponds to the allowed range of shadow diameter, spin, and circularity deviation for M$87^{\ast}$ as reported by EHT. On the other hand, by considering a minimum shadow diameter of $9.5$, we were able to place constraints on the radial and latitudinal plasma parameters, with maximum values $k_r=4.75$ and $k_{\theta}=85.5$, respectively. These constraints provide valuable insights into the physical properties of the wormhole and the plasma surrounding it. By examining the maximum and minimum shadow diameters, we can better understand the range of possible sizes for the throat and the corresponding plasma parameters that are consistent with the observed shadows in the case of M$87^{\ast}$. 

In our future research, we intend to investigate the impact of plasma on the shadow of a Kerr black hole. Additionally, we plan to compare the findings from the study of wormhole shadows to those of black hole shadows. This comparative analysis will provide further insights and potentially help discern whether M$87^{\ast}$ is more likely to be a black hole or a wormhole. By delving into these investigations, we hope to contribute to the ongoing understanding of M$87^{\ast}$ and its intriguing nature, paving the way for deeper insights into the astrophysical phenomena occurring in the vicinity of these enigmatic cosmic objects.

\section*{Acknowledgement}

The work of SC is supported by Mathematical Research Impact Centric Support (MATRICS) from the Science and Engineering Research Board (SERB) of India
through grant MTR/2022/000318.

\bibliography{references}
\newpage
\appendix
\section{Calculations for the deflection angle }
\label{Appendix A}
Let us expand equation \ref{6.10} for the homogeneous plasma with low plasma density in the context of a slow-rotating wormhole, assuming $r_0/R<1$.
\begin{equation}
    \bar{\alpha}  = \int_{-\infty}^{\infty} \Phi_0 dr+\int_{-\infty}^{\infty} \Phi_1 dr+ \int_{-\infty}^{\infty} \Phi_2 dr ,
\end{equation}
where the integrand is given as,
\begin{align}
\begin{split}
\Phi_0 &= \frac{2R}{r\sqrt{r^2-R^2}}, \\
\Phi_1 &= \frac{r_0 \left(2 r^2+r R+R^2\right)}{r^2 (r+R) \sqrt{r^2-R^2}} \\
&\quad +a\frac{2 r_0 \left(4 r^2+7 r R+9 R^2\right)+4 r R (r+R)}{r R^2 (r+R)^2 \sqrt{r^2-R^2}}, \\
\Phi_2 &= \Bigg(\frac{2 r_0}{(r+R) \sqrt{r^2-R^2}}\\
&\quad +a \frac{r_0 \left(4 r^2+3 r R+17 R^2\right)+2 r R (r+R)}{r R^2 (r+R)^2 \sqrt{r^2-R^2}}\Bigg)\frac{k_0}{R^2}.
\end{split}
\end{align}
and, upon solving while neglecting higher-order terms, we get the deflection angle by Teo wormhole in uniform plasma space-time as
\begin{equation}
     \bar{\alpha} = \pi+\left(\frac{3r_0}{R}+\frac{4a}{R^2}+\left(\frac{2r_0 k_0}{R}+\frac{2a k_0}{R^2}\right)\right).
\end{equation}
Similarly, in the case of non-homogeneous plasma distribution let us again expand equation \ref{6.10} in the context of a slow-rotating wormhole by assuming $r_0/R<1$ and $\omega_P(R)^2/\omega_o^2<1$,
\begin{equation}
    \bar{\alpha}  = \int_{-\infty}^{\infty} \psi_0 dr+ \int_{-\infty}^{\infty} \psi_1 dr+ \int_{-\infty}^{\infty} \psi_2 dr,
\end{equation}
where the integrand is given as,
\begin{align}
\begin{split}
\psi_0 &= \frac{2R}{r\sqrt{r^2-R^2}}, \\
\psi_1 &= \frac{r_0 \left(2 r^2+r R+R^2\right)}{r^2 (r+R) \sqrt{r^2-R^2}} \\
&\quad + a\frac{2 r_0 \left(4 r^2+7 r R+9 R^2\right)+4 r R (r+R)}{r R^2 (r+R)^2 \sqrt{r^2-R^2}},\\
\psi_2 &= \Bigg(\frac{r_0 \left(2 r^2+3 r R-R^2\right)-2 r R (r+R)}{2 r^2 (r+R) \sqrt{r^2-R^2}}\\
&\quad -a\frac{2 \left(r_0 \left(4 r^2+3 r R+5 R^2\right)+2 r R (r+R)\right)}{r^3 (r+R)^2 \sqrt{r^2-R^2}}\Bigg)\frac{k_0}{R^2}.
\end{split}
\end{align}

and, upon solving while neglecting higher-order terms, we get the deflection angle by Teo wormhole in radial plasma space-time as

\begin{equation}
\begin{split}
    \bar{\alpha}= \pi+\Bigg(\frac{3r_0}{R}+\frac{4a}{R^2}+\frac{ r_0 a (9\pi-14)}{R^3}\\+\frac{k_r}{2R^2}\left(\frac{r_0 (2\pi-3)} {R}-\pi \right)\Bigg).
\end{split}
\end{equation}

\end{document}